\documentclass[a4paper,11pt]{amsart}

\usepackage{graphicx}
\usepackage{a4wide}
\usepackage[dvipsnames]{xcolor}
\usepackage{epstopdf,epsfig}
\usepackage{amsaddr,amsmath,amsfonts,amssymb,amsgen,amsbsy,dsfont}
\usepackage{mathtools,bm}
\usepackage{subfig}
\usepackage{upgreek}
\usepackage{hyperref}
\usepackage{gensymb}
\usepackage[natbibapa]{apacite}

\DeclareMathAlphabet{\pazocal}{OMS}{zplm}{m}{n}

\newcommand{\crit}[1]{{#1}_\mathrm{c}}
\newcommand{\rf}[1]{(\ref{#1})}
\newcommand{\eqdef}{\vcentcolon=}

\newcommand{\Ro}{\mathrm{Ro}}
\newcommand{\Ek}{\mathrm{Ek}}

\DeclareMathOperator{\Hess}{Hess}
\DeclareMathOperator{\Tr}{tr}

\newcommand{\ud}{\mathrm{d}}

\newcommand{\ue}{\mathrm{e}}
\newcommand{\ui}{\mathrm{i}}

\newcommand{\vu}{\boldsymbol{u}}                    
\newcommand{\nab}{\boldsymbol{\nabla}}              
\newcommand{\scal}{\boldsymbol{\cdot}}              

\begin{document}

\title[On the stability of viscous Riemann ellipsoids]{On the stability of viscous Riemann ellipsoids}

\author[J. Labarbe]{Joris Labarbe}
\address{Universit\'e C\^ote d'Azur, CNRS UMR 7351,  Laboratoire J.~A. Dieudonn\'e, 
Parc Valrose, 06108 Nice cedex 2, France}
\email{joris.LABARBE@univ-cotedazur.fr}


\begin{abstract}
The present study investigates the linear stability of Riemann ellipsoids in 
both the inviscid limit and in the presence of weak viscosity. In the inviscid 
regime, we derive a generalised Poincar\'e equation governing small fluid 
oscillations and construct a family of polynomial solutions that extends the 
classical results of Cartan to flows with a uniform strain field. This formulation 
provides an analytic dispersion relation for three-dimensional ellipsoidal 
disturbances and remains computationally efficient at arbitrary harmonic degree, in 
contrast to the virial tensor method or to short-wavelength (WKB) approximations. 
The viscous effects are incorporated through a boundary-layer analysis based on Prandtl’s 
theory, leading to first-order viscous corrections to the inviscid spectrum and allowing 
a systematic investigation of viscosity-driven instabilities. Stability diagrams 
are presented over the space of admissible Riemann ellipsoids, illustrating the 
roles of rotation, internal strain, and diffusion, with implications 
for rotating shear flows in geophysical and astrophysical contexts.
\end{abstract}

\maketitle

\section{Introduction} \label{sec_intro}

Celestial mechanics and the laws of planetary motion rank among the oldest 
and most prominent areas of research in classical physics. Historically, 
the formulation of Newtonian mechanics together with the universal law of 
gravitation provided a unified and quantitative framework within which such 
problems could be addressed rigorously. Since then, sustained interest has 
developed in the study of the equilibrium and stability of self-gravitating 
fluids, motivated both by fundamental theoretical considerations and by 
applications in astrophysics and geophysics. In his treatise of fluxions, 
\cite{Maclaurin} identified the only stable axisymmetric family of equilibrium 
figures corresponding to rigidly rotating, homogeneous fluid bodies in a 
gravitational field. These configurations, now known as Maclaurin spheroids, 
were subsequently investigated by numerous authors as simplified models of 
isolated rotating stars (or planets) subject to the Coriolis effect and 
centrifugal acceleration. Over a century later, \cite{M1842} and \cite{L1851}, 
employing distinct analytical approaches, demonstrated that the Maclaurin 
sequence undergoes a bifurcation at a finite eccentricity towards a 
non-axisymmetric equilibrium, namely the Jacobi ellipsoid \citep{J1839}. 
These results were placed on a rigorous footing by \cite{P85}, who interpreted 
the departure from the initial state within the modern framework of 
dynamical systems theory and bifurcation analysis, thereby identifying one 
of the earliest paradigmatic example of spontaneous symmetry breaking in 
continuum mechanics. In the same monograph, Poincaré derived a differential 
equation governing the small oscillations of a rotating, incompressible, 
self-gravitating, inviscid, and strainless mass of fluid. This formulation 
stimulated the work of \cite{B89}, who analysed the normal modes of Maclaurin 
spheroids through solutions of what is now referred to as the Poincaré equation. 
Several decades later, \cite{C22} revisited the problem and rigorously 
identified the unique family of polynomial solutions satisfying this equation. 
In the absence of self-gravitational effects, \cite{G65} showed that the associated 
Poincaré operator is self-adjoint and possesses a bounded real spectrum. He 
further raised the question of the completeness of the corresponding inertial 
modes, a fundamental issue for the modal decomposition of rotating flows, which 
was resolved affirmatively only much later for arbitrary bounded ellipsoidal 
domains \citep{BR17}.

As shown by \cite{D1861} in a seminal contribution, to every equilibrium figure 
whose velocity field is generated by a linear mapping of the position vector 
there exists an adjoint configuration whose velocity field is determined by the 
transpose of this linear map. Remarkably, the original and adjoint configurations 
possess the same ellipsoidal shape and both satisfy the conditions of gravitational 
equilibrium. Owing to the linear dependence of the velocity field on the spatial 
coordinates, these flows are characterised, in the rotating frame, by a uniform 
rate of strain and a uniform vorticity. Building on Dedekind’s theorem, \cite{R60} 
provided a complete classification of all admissible equilibrium configurations 
with respect to Dedekind's assumptions. In particular, Riemann identified three 
distinct families of non-axisymmetric ellipsoids obtained by superposing solid-body 
rotation with internal shear motions of uniform vorticity. Among these, the 
so-called S-type ellipsoids --- whose vorticity vector is aligned with the rotation 
axis and therefore with the Coriolis force --- are of particular relevance for physical 
applications. Despite their potential importance for modelling internal differential 
motions in planets and stars, and their role as the simplest exact solutions 
incorporating both rotation and strain, these equilibria were largely overlooked 
for many decades. A systematic linear stability analysis of Riemann ellipsoids 
was eventually undertaken by \cite{C65,C66}, who employed the virial tensor method 
developed and applied extensively in related contexts \citep{EFE}. Nevertheless, 
a comprehensive linear stability theory for Riemann ellipsoids, valid for harmonic 
perturbations of arbitrary order and comparable to that established for Maclaurin 
spheroids \citep{C79}, has yet to be developed.

In the specific context of rigidly rotating spheroids, \cite{R60} identified the 
onset of a dynamical instability at a finite eccentricity, corresponding to disk-like 
configurations. The associated critical threshold describes a Hamilton--Hopf 
bifurcation, which can be interpreted within Krein theory as the collision of two 
eigenvalues with opposite Krein signatures, or energy signs \citep{Labarbe}. Such 
instabilities are nowadays recognised as generic features of non-dissipative systems 
with specific symmetries, and similar phenomena arise well beyond the setting of 
Maclaurin spheroids, appearing in a broad class of out-of-equilibrium Hamiltonian 
systems \citep{Kirillov}. On the basis of Dirichlet’s principle, \cite{R60} further 
argued that his non-homogeneous ellipsoids are stable for almost all admissible 
shapes and rotation rates. This conclusion was subsequently challenged by \cite{C65,C66}, 
who demonstrated that a wide range of self-gravitating configurations are unstable 
with respect to second- and third-order harmonic perturbations, thereby contradicting 
Riemann’s original claim. These instability results were later corroborated by \cite{LL96} 
using a short-wavelength, geometrical-optics (WKB) approximation of the linearised 
equations of motion. It is now understood that the discrepancy between the conclusions 
of Riemann and those of Chandrasekhar arises from an incorrect application of 
Dirichlet’s principle, since the governing equations of motion were not casted into an 
admissible Lagrangian form \citep{MLB09}.

Instability is an inherent feature of inviscid fluid systems with elliptic streamlines. 
From a physical standpoint, such instability arises from resonant interactions between 
the normal modes of oscillation and an imposed planar strain field \citep{K02}. Riemann 
ellipsoids may therefore be regarded as among the earliest examples of this phenomenon, 
although their connection to elliptic instability remained naturally unrecognised for many 
decades. Subsequent studies demonstrated that a uniform strain field can excite Kelvin 
modes in the form of helical waves, in axisymmetric vortices \citep{MS75} as well as in unbounded 
vortex flows \citep{P86,B86}. These works established that elliptic instability generates 
local enstrophy along the vortex axis through vortex stretching, thereby highlighting its 
fundamental three-dimensional character. Given the central role of enstrophy production 
in turbulence, this mechanism has also been extensively investigated in the context of shear-flow 
transition \citep{BOH88}. When viscosity is incorporated into the stability analysis, it 
introduces a damping contribution that weakens growth rates without suppressing the 
underlying instability mechanism \citep{LS87}. However, because dissipation alters the 
global energy balance, it can lead to significant modifications of the nonlinear saturation 
and long-term dynamics relative to the conservative case.

Energy dissipation, although often regarded as a stabilising phenomenon, can in certain 
circumstances act as a destabilising mechanism. This counterintuitive behaviour was first 
identified in the classical study of \cite{KelvinI}, in the context of Maclaurin 
spheroids endowed with arbitrarily small viscosity. Nearly a century later, \cite{RS63} 
rigorously established the existence of this secular instability using asymptotic methods 
and boundary-layer theory. They actually showed that sectoral modes of rigidly rotating stars 
exhibit exponential growth once the eccentricity exceeds a critical value, which coincides 
with the bifurcation point from the Maclaurin to the Jacobi sequence identified by 
\cite{M1842} and \cite{L1851}. This secular instability is now understood as a prototypical 
example of dissipation-induced instability and is commonly attributed to the destruction 
of gyroscopic stabilisation by diffusion \citep{BKMR94,Kirillov,Labarbe}. With 
regard to this phenomenon, and as already noted by \cite{LL96}, a systematic treatment 
of viscosity-induced instabilities in the context of Riemann ellipsoids remains 
conspicuously absent from the literature.

In a further seminal contribution, \cite{C70} demonstrated the existence of 
dissipation-induced instabilities in Maclaurin spheroids arising from diffusive effects 
associated with gravitational radiation reaction, within a post-Newtonian framework. 
One of the principal implications of this result is that sufficiently massive, isolated 
rotating stars may emit gravitational waves as a consequence of the 
Chandrasekhar–Friedman–Schutz (CFS) instability \cite{C70,FS78a,FS78b}. This mechanism 
has since become a promising objective in gravitational-wave astrophysics, particularly 
following the direct detection of gravitational waves from compact binary mergers by 
ground-based interferometers \citep{A16}. Subsequently, \cite{C79} generalised these 
results to harmonic perturbations of arbitrary order and demonstrated that the secular 
modes destabilised by viscosity are distinct from those driven unstable by gravitational 
radiation reaction. To date, no comparable comprehensive analysis has been carried out 
for Riemann ellipsoids, beyond the inviscid investigations of \cite{C65,C66} and \cite{LL96}.

The present work aims to fill the aforementioned gaps by developing a unified and tractable 
linear stability theory for Riemann ellipsoids, valid both in the inviscid limit and in the 
presence of weak viscosity. In the inviscid setting, we introduce a generalised Poincaré 
equation governing small-amplitude oscillations about the equilibrium state and construct 
an associated family of polynomial solutions that encompasses, as a special case, the 
classical solutions obtained by Cartan. This formulation yields an explicit analytic 
dispersion relation for inviscid Riemann ellipsoids and allows stability properties to 
be computed efficiently for ellipsoidal harmonics of arbitrary order. In contrast to the 
virial tensor method introduced by \cite{EFE}, the present approach remains 
computationally efficient at high harmonic degree and provides direct spectral access to 
the full mode structure. Moreover, unlike the short-wavelength (WKB) approximation employed 
by \cite{LL96}, the analysis applies uniformly across all wavelengths. In the presence of 
viscosity, we derive a corresponding analytic dispersion relation by means of a boundary-layer 
analysis based on Prandtl’s theory \citep{LandauLifshitz}, yielding first-order viscous corrections to 
the inviscid spectrum and enabling a systematic investigation of viscosity-driven effects. 

The structure of the article is as follows. In Section~\ref{sec_pb}, we introduce the 
governing equations of motion together with the relevant boundary conditions, and describe 
the inviscid equilibrium state as well as the linearised equations about this steady 
solution. Section~\ref{sec_inv} presents the derivation of the equation governing the
normal modes of oscillations and its general polynomial solution. In Section~\ref{sec_vis}, 
we develop the small-viscosity approximation and obtain the first-order viscous corrections 
to the dispersion relation. Selected stability results across the relevant parameter space 
are then presented. Finally, Section~\ref{sec_conclu} discusses the physical interpretation 
of these findings in the broader contexts of geophysics and astrophysics, highlighting 
their implications for rotating fluid bodies with internal motion.

\section{Formulation of the problem} \label{sec_pb}

\subsection{Equations of motion and equilibrium state}

\begin{figure}[t!]
\includegraphics[width=.65\textwidth]{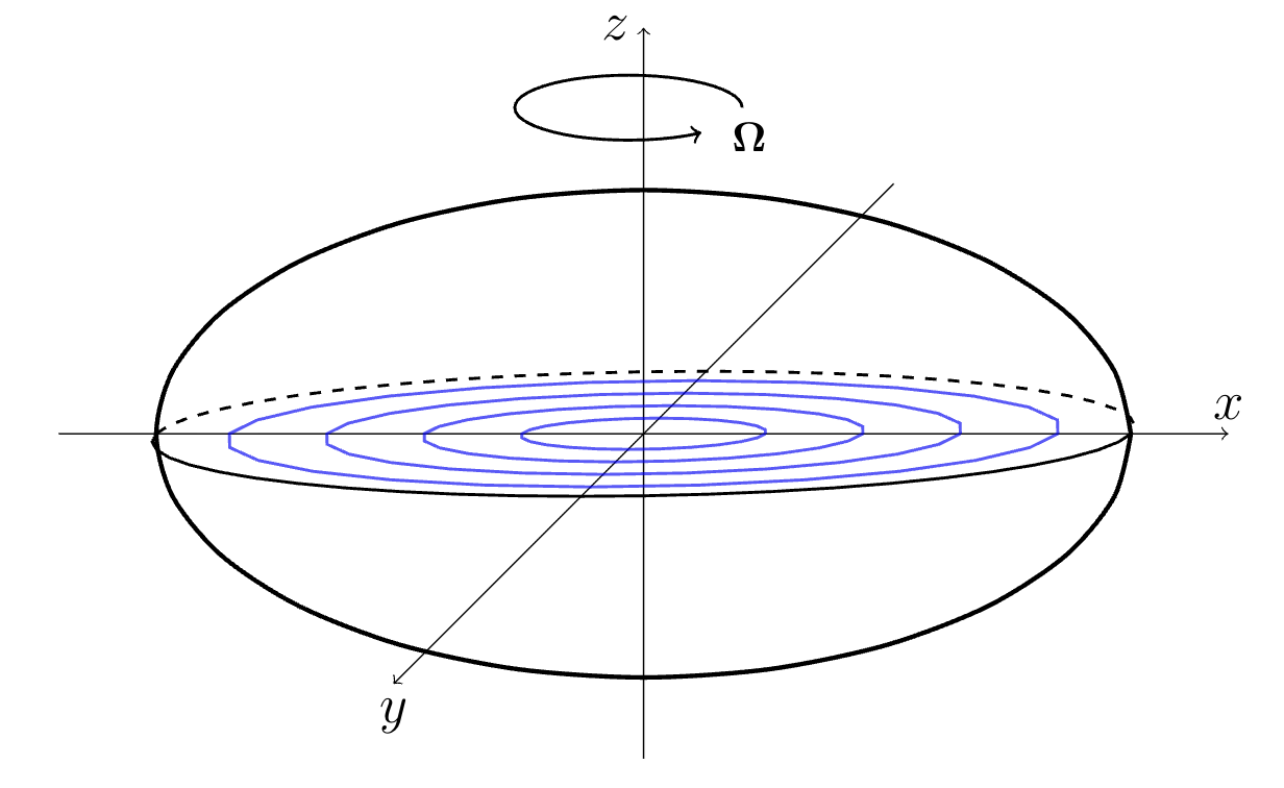}
\caption{Sketch of a rigidly rotating and self-gravitating Riemann ellipsoid 
with uniform strain at its interior. In blue are depicted the elliptic streamlines.
\label{riem_ell}}
\end{figure}

Riemann ellipsoids (of S-type family, as exclusively restricted in this work) are equilibrium 
configurations of self-gravitating, homogeneous and incompressible masses of fluid. They are 
characterised by a combination of solid-body rotation and uniform internal vorticity, both 
aligned with a fixed vertical axis. The rigid rotation of the ellipsoid is described by the 
angular velocity $\Omega$, gravitational effects are modelled through the Newtonian potential 
$\Phi$, and the constant vorticity in the co-rotating frame is denoted by $\zeta$. The assumption 
of uniform vorticity implies that the associated velocity field depends linearly on the spatial 
coordinates, according to Dedekind's theorem for adjoint ellipsoidal configurations \citep{D1861}.
A schematic representation of an arbitrary Riemann ellipsoid is depicted in Figure~\ref{riem_ell}.

The fluid motion is described either in Cartesian coordinates $(x,y,z)$, with respect to an 
orthonormal basis $(\bm{e}_x,\bm{e}_y,\bm{e}_z)$, or in ellipsoidal coordinates, better suited 
to the geometry of the equilibrium figure. The latter coordinate system, together with 
a selection of geometrical relations, is presented in Appendix~\ref{app1}. Without loss
of generality, we assume the semi-axes of the ellipsoid to respect $0 < a_3 \leq a_2 \leq a_1 < \infty$. 
We denote the characteristic time frequency by $(\pi\rho G)^{1/2}$ and the unit length 
by $a_1$, the largest semi-axis of the ellipsoid. Thus, the zero level set of the function
\begin{equation}
\label{ellipsoid}
F(x,y,z) = x^2 + \frac{y^2}{\Gamma^2} + \frac{z^2}{\Xi^2} - 1 ,
\end{equation}
denotes the surface of the Riemann ellipsoid. In the latter, $\Gamma\eqdef a_2/a_1$ and 
$\Xi\eqdef a_3/a_1$ are the semi-axes aspect ratios.

The governing equations for the velocity and pressure fields are given by 
the incompressible Navier--Stokes equations. When written in dimensionless 
form and in the non-inertial frame of reference, they read
\begin{subequations}
\label{eom}
\begin{align}
\frac{\partial\vu}{\partial t} + \vu\scal\nab\vu + 2\Ro(\bm{e}_z\times\vu) +
Ro^2 \bm{e}_z \times ( \bm{e}_z \times \bm{r} ) &= \nab ( \Phi  - p ) + \Ro \Ek \nab^2 \vu , \\
\nab\scal\vu &= 0 ,
\end{align}
\end{subequations}
where $\vu(\bm{r},t)$ denotes the velocity field, $p(\bm{r},t)$ the pressure and $\bm{r}=
x\bm{e}_x+y\bm{e}_y+z\bm{e}_z$ the Cartesian position vector.
The Newtonian gravitational potential $\Phi$ is given by
\begin{equation}
\label{newton}
\Phi(\bm{r},t) = \frac{1}{\pi} 
\int_{\mathcal{D}_t} \frac{\ud\bm{r}'}{\vert\bm{r}-\bm{r}'\vert} ,
\end{equation}
with the integration performed over the time-dependent fluid domain $\mathcal{D}_t$.

In the above expressions, $\Ro$ and $\Ek$ are the dimensionless Rossby and Ekman 
numbers, respectively, defined as
\begin{equation}
\Ro \eqdef \frac{\Omega}{(\pi \rho G)^{1/2}}, \quad
\Ek \eqdef \frac{\nu}{a_1^2 \Omega} ,
\end{equation}
where $\rho$ is the constant fluid density, $G$ the gravitational constant, and $\nu$
the kinematic viscosity.

We introduce the Cauchy stress tensor of the fluid
\begin{equation}
\sigma_{ij} \eqdef - p \delta_{ij} + 2\Ro\Ek \varepsilon_{ij} = 
- p \delta_{ij} + \Ro\Ek \left( \partial_i u_j + \partial_j u_i \right) ,
\end{equation}
where $\delta$ denotes the Kronecker delta function and $(i,j)$ label the components in 
the chosen coordinate system (the conversion of tensor components between two curvilinear bases is given in 
Appendix~\ref{app2}). The surface of the ellipsoid being a free surface, we impose a dynamic 
no-stress boundary condition. Defining the outward unit normal and tangential vectors by 
$\bm{n}$ and $\bm{t}$, this condition reads $\bm{n}\scal(\sigma\scal\bm{n}) = 0$ and 
$\bm{t}\scal(\sigma\scal\bm{n}) = 0$ (assuming vacuum outside the ellipsoid). In addition, 
the velocity field is required to satisfy a kinematic boundary condition enforcing continuity 
at the moving free surface.

We note that for $\Ek\to 0$ (i.e. for the inviscid limit $\nu\to0$), the no-shear stress 
condition becomes redundant and thus must be discarded. In the viscous case, however, 
consistency of the solution requires the presence of a thin boundary-layer adjacent to 
the inner surface of the ellipsoid, within which viscous effects remain significant.

The steady-state velocity field of Riemann ellipsoids is linear in the Cartesian coordinates.
Explicitly, they are given by $\vu_0 = \bm{S} \bm{r}$ \citep{R60}, where
\begin{equation}
\label{S}
\bm{S} \eqdef 
\begin{pmatrix}
0 & Q_1 & 0 \\
Q_2 & 0 & 0 \\
0 & 0 & 0
\end{pmatrix} , \quad \textrm{with} \quad
Q_1 \eqdef -\frac{\zeta}{1+\Gamma^2} \quad \textrm{and} \quad
Q_2 \eqdef \frac{\zeta\Gamma^2}{1+\Gamma^2} .
\end{equation}
We note that this solution is harmonic, i.e. it satisfies $\nab^2\vu_0\equiv \bm{0}$. 
Therefore, the Riemann velocity field satisfies equally the Euler and the Navier--Stokes 
equations. Accordingly, the steady pressure and symmetric part of the velocity tensor 
are given by
\begin{equation}
p_0 = - P_0 F(\bm{r}) , \quad \nab\vu_0 + (\nab\vu_0)^T = \bm{S} + \bm{S}^T = 
( Q_1 + Q_2 )
\begin{pmatrix}
0 & 1 & 0 \\
1 & 0 & 0 \\
0 & 0 & 0
\end{pmatrix} ,
\end{equation}
where $F(x,y,z)$ is the function defined in \rf{ellipsoid} and $P_0>0$ is recovered from 
the equations of hydrostatic balance. The transpose operator is denoted by the 
superscript $T$. We note that the strain-rate tensor vanishes identically for $\Gamma=1$ 
(i.e. $a_1=a_2$) or $\zeta=0$, which corresponds to the Maclaurin and Jacobi families of rigidly 
rotating ellipsoids. The latter configuration and its stability have been notably studied in \cite{C62}.

To characterise admissible equilibrium configurations of Riemann ellipsoids, we follow the 
classical construction of \cite{C65} and introduce the dimensionless ratio of internal vorticity 
to rigid rotation, $f\eqdef\zeta/\Ro$. Fixing the value of $f$ defines a \textit{Riemann sequence} of 
equilibria parameterised by the ellipsoidal geometry. In particular, the limiting cases $f=0$ 
and $f\to\pm\infty$ correspond to the Jacobi and Dedekind sequences, respectively, representing 
purely rotational and purely internal-flow-supported equilibria. For such configurations to 
satisfy gravitational equilibrium, Chandrasekhar derived the coupled algebraic conditions
\begin{equation}
\begin{multlined}
\label{adm_Xi}
\Gamma^2 \Big[ A_1 - A_2 + \frac{\Xi^2}{\Gamma^2} \big( 1 - \Gamma^2 \big) A_3 \Big] f^2
+ 2 \big( 1 + \Gamma^2 \big) \big( A_1 - A_2 \Gamma^2 \big) f \\
+ \big( 1 + \Gamma^2 \big)^2 \Big[ A_1 - A_2 + \frac{\Xi^2}{\Gamma^2} \big( 1 - \Gamma^2 \big) A_3 \Big] = 0 ,
\end{multlined}
\end{equation}
and
\begin{equation}
\label{adm_Ro}
\Ro^2 = \frac{2B_{12}}{1 + f^2\Gamma^2/(1+\Gamma^2)^2} ,
\end{equation}
where the coefficients $A_i$ and $B_{ij}$ are the classical index symbols 
defined in \citep{EFE}. For a prescribed value of $f$, solving 
\eqref{adm_Xi} determines the admissible ellipsoidal geometry (through 
$\Gamma$ or $\Xi$), while \eqref{adm_Ro} subsequently fixes the 
corresponding Rossby number. Fixing $f=0$ and $\Gamma=1$, the Rossby
number of Maclaurin spheroids is expressed solely in terms of the 
eccentricity $e\eqdef d_\lambda$ \citep{EFE,Labarbe}.

\subsection{Linear stability theory}

We restrict this work to slightly perturbed Riemann ellipsoids for the linear theory to
remain valid. We denote by $\bm{\xi}(\bm{r},t)$ the Lagrangian displacement, which refers 
to a fluid parcel which would have been at time $t$ and position $\bm{r}$ in the 
unperturbed configuration. Accordingly, we introduce the Lagrangian difference operator as
\begin{equation}
\label{diff_lag}
\Delta f(\bm{r},t) \eqdef f(\bm{r}+\bm{\xi}(\bm{r},t),t) - f_0(\bm{r},t) ,
\end{equation}
where $f$ refers to an arbitrary quantity evaluated in the perturbed configuration, 
whereas $f_0$ refers to the value of the same quantity but for the unperturbed flow ---
the \textit{ghostly flow} solution in the terminology of \cite{LBO67}. 
The Eulerian change operator is recovered by setting $\bm{\xi}\equiv0$ in the 
definition \rf{diff_lag}, reading
\begin{equation}
\label{diff_eul}
\delta f(\bm{r},t) \eqdef f(\bm{r},t) - f_0(\bm{r},t) .
\end{equation}

It is worth mentioning that \cite{FS78a,FS78b} introduced a modified operator in 
their work on Lagrangian perturbation theory. The operator they define involves the 
Lie derivative with respect to the Lagrangian displacement and allow for a whole 
description in Lagrangian frame, without need to introduce a mixed Eulerian--Lagrangian 
formalism. Moreover, adopting the description of Friedman and Schutz eliminates the 
gauge freedom in the canonical Hamiltonian of the system and thus yields meaningful 
stability criteria. However, this Lagrangian picture is not suited to study viscous 
flows as the classical Laplacian would become a Laplace--Beltrami operator for the 
disturbed metric, rendering the problem highly nonlinear. Therefore, our linear theory
is solely based on a mixed Lagrangian--Eulerian formulation to allow for viscosity
to be taken into account, as done by \cite{RS63}.

We apply operator \rf{diff_lag} on \rf{eom}--\rf{newton} and expand the solution
in Taylor series for arbitrarily small displacements. 
We recover the linearised equations of motion in a mixed Eulerian--Lagrangian description
by retaining only first-order terms in disturbances 
\begin{subequations}
\label{leom}
\begin{align}
\Big( \frac{\partial}{\partial t} - \Ro\Ek\nab^2 + \vu_0\scal\nab + \bm{S} + 
2\Ro\bm{e}_z\times \Big) \delta\vu &= \nab \big( \delta\Phi -  \delta p \big) \label{leom1} , \\
\nab \scal \delta\vu &= 0 , \label{leom2} \\
\delta\Phi(\bm{r},t) &= \frac{1}{\pi} \int_{\partial\mathcal{D}_0} 
\frac{\bm{\xi}(\bm{r}',t)\scal\bm{n}}{\vert\bm{r}-\bm{r}'\vert} \ud\bm{r}' ,
\end{align}
\end{subequations}
where $\partial\pazocal{D}_0$ is the boundary of the equilibrium ellipsoid, i.e. the
surface of equation $\lambda=1$. As described in Appendix \ref{app1}, the outward 
unit normal vector is given by $\bm{n}\eqdef\bm{e}_\lambda$.

Applying the same procedure as before, the linearised boundary conditions are written 
on the unperturbed free surface and in ellipsoidal coordinates (cf. \cite{C79} for details) as
\begin{equation}
\label{lbc}
\delta p + \bm{\xi}\scal\nab p_0 - 2\Ro\Ek\varepsilon_{\lambda\lambda}  = 0 , \quad
\varepsilon_{\lambda\mu} = 0 , \quad
\varepsilon_{\lambda\varphi} = 0 ,
\end{equation}
supplemented with kinematic condition.
Despite the flow having internal motion with non-vanishing components 
of the stress tensor, their contributions to the linearised boundary conditions
are not present since they remain uniform at all time and thus, are of the same order 
than the background state. Moreover, the disturbance fields are decomposed
as $\delta\vu \eqdef \delta\vu^{(0)} + \delta\vu^{(\nu)}$, 
$\delta p \eqdef \delta p^{(0)} + \delta p^{(\nu)}$ and 
$\bm{\xi} \eqdef \bm{\xi}^{(0)} + \bm{\xi}^{(\nu)}$, where the superscripts
denote whether these fields are solutions to the inviscid or the viscous equations.

In the following, we present an exhaustive analysis of the inviscid configuration, 
where we obtain a dispersion relation for arbitrary harmonic degree, extending the 
existing results of \cite{C65,C66} who used the tensor virial method. Subsequently, 
we present a procedure to solve the viscous problem by means of a boundary-layer theory, 
as done by \cite{RS63} and \cite{C79} in the context of Maclaurin spheroids. Without 
loss of generality, we assume the disturbances to have harmonic motions of the form 
$\exp{(-2\ui\omega t)}$. The real and imaginary parts of the eigenfrequency $\omega$ 
correspond to, respectively, the growth rate (in magnitude) and the frequency of the disturbances.

\section{Inviscid Riemann ellipsoids} \label{sec_inv}

The boundary condition for the inviscid problem is obtained by taking the limit
$\Ek \to 0$ in \rf{lbc}, which yields
\begin{equation}
\label{lbc_inv}
\big[ \delta p^{(0)} \big]_{\partial\mathcal{D}_0} - 2 P_0 
\big[ h_\lambda \xi_\lambda^{(0)} \big]_{\partial\mathcal{D}_0} = 0 ,
\end{equation}
where we have used the relation 
$\nab p_0 = -\vert\nab p_0\vert \bm{n} = -2 P_0 h_\lambda \bm{e}_\lambda$.
Our objective is to evaluate the different terms in \rf{lbc_inv} in order to derive 
a dispersion relation for the eigenfrequency $\omega$, and thereby assess the linear 
stability of the system for admissible configurations.

\subsection{Harmonic expansion of disturbance fields}

The product of the normal Lagrangian displacement and the corresponding metric scale factor,
evaluated at the equilibrium boundary $\partial\mathcal{D}_0$, is expanded in terms of
surface ellipsoidal harmonics --- whose definitions are given in Appendix~\ref{app3} --- as
\begin{equation}
\label{xi}
\big[ h_\lambda \xi_\lambda^{(0)} \big]_{\partial\mathcal{D}_0} = \widehat{\xi}_{nm}
\mathbb{S}_n^m(\mu,\varphi) \ue^{-2\ui\omega t} ,
\end{equation}
The corresponding expansion coefficients are given by
\begin{equation}
\label{cte_xi}
\widehat{\xi}_{nm} = \frac{\ue^{2\ui\omega t}}{\varpi_n^m} \int 
\big[ h_\lambda \xi_\lambda^{(0)} \big]_{\partial\mathcal{D}_0} \mathbb{S}_n^m(\mu,\varphi) \ud S ,
\end{equation}
where integration is performed over the surface of the equilibrium ellipsoid.
Unless stated otherwise, all integrals for the surface element $\ud S$ appearing 
in the remainder of this article are taken over this same geometry.

Using the definition of the Lagrangian displacement,
\begin{equation}
\label{xi_def}
\delta \vu^{(0)} = \frac{\partial\bm{\xi}^{(0)}}{\partial t} = -2\ui\omega\bm{\xi}^{(0)} ,
\end{equation}
together with expression \rf{vec_conv}, the coefficients \rf{cte_xi} can be expressed
in terms of the Cartesian components of the inviscid velocity perturbation as
\begin{equation}
\label{cte_xi_u}
\widehat{\xi}_{nm} = \frac{\ui\ue^{2\ui\omega t}}{2\omega\varpi_n^m} \int
\Big[ \frac{\partial x}{\partial\lambda} \delta u_x^{(0)} + \frac{\partial y}{\partial\lambda} \delta u_y^{(0)}
+ \frac{\partial z}{\partial\lambda} \delta u_z^{(0)} \Big]_{\partial\mathcal{D}_0} \mathbb{S}_n^m(\mu,\varphi) \ud S .
\end{equation}

Using the expansion \rf{ell_expand} and exploiting the symmetry of the problem,
the perturbation to the Newtonian gravitational potential can be expressed
throughout space as \citep{EFE}
\begin{equation}
\label{Phi0}
\delta\Phi^{(0)} (\lambda,\mu,\varphi,t) = 
\begin{cases}
\begin{aligned}
\widehat{\Phi}_{nm} \dfrac{E_n^m(\lambda)}{E_n^m(1)} \mathbb{S}_n^m(\mu,\varphi) \ue^{-2\ui\omega t} , 
& \quad d_\mu \leq \lambda \leq 1 , \\[.5ex]
\widehat{\Phi}_{nm} \dfrac{F_n^m(\lambda)}{F_n^m(1)} \mathbb{S}_n^m(\mu,\varphi) \ue^{-2\ui\omega t} , 
& \quad 1 \leq \lambda < \infty ,
\end{aligned}
\end{cases}
\end{equation}
where the interior and exterior ellipsoidal harmonics are introduced in Appendix~\ref{app3}.

The expansion coefficients are obtained by applying Gauss’ theorem at the
unperturbed boundary $\partial\mathcal{D}_0$, namely
\begin{equation}
\label{gauss_th}
\big[ \bm{n}\scal\nab\delta\Phi^{(0)} \big]_{\partial\mathcal{D}_0^+} -
\big[ \bm{n}\scal\nab\delta\Phi^{(0)} \big]_{\partial\mathcal{D}_0^-} =
- 4 \big[ \bm{n} \scal \bm{\xi}^{(0)} \big]_{\partial\mathcal{D}_0} ,
\end{equation}
where $\partial\mathcal{D}_0^\pm$ denote the exterior and interior sides of the boundary,
respectively. In ellipsoidal coordinates, this condition can be rewritten such that
the right-hand side directly involves the quantity defined in \rf{xi}. Explicitly,
\begin{equation}
\label{gauss_th2}
\Big[ \frac{\partial\delta\Phi^{(0)}}{\partial\lambda} \Big]_{\partial\mathcal{D}_0^+} -
\Big[ \frac{\partial\delta\Phi^{(0)}}{\partial\lambda} \Big]_{\partial\mathcal{D}_0^-} =
- 4 \big[ h_\lambda \xi_{\lambda}^{(0)} \big]_{\partial\mathcal{D}_0} =
- 4 \widehat{\xi}_{nm} \mathbb{S}_n^m \ue^{-2\ui\omega t} .
\end{equation}

Substituting \rf{Phi0} into \rf{gauss_th2} and making use of the Wronskian relation
\rf{Wronsk}, we obtain
\begin{align}
\label{delta_phi0}
\big[ \delta\Phi^{(0)} \big]_{\partial\mathcal{D}_0} 
&= 4 \widehat{\xi}_{nm} \pazocal{W}_n(1)^{-1} E_n^m(1) F_n^m(1) 
\mathbb{S}_n^m(\mu,\varphi) \ue^{-2\ui\omega t} \\
&= \frac{4 \widehat{\xi}_{nm}}{2n+1} \sqrt{(1 - d_\mu^2)(1 - d_\varphi^2)} 
E_n^m(1) F_n^m(1) \mathbb{S}_n^m(\mu,\varphi) \ue^{-2\ui\omega t} .
\end{align}

To complete the inviscid boundary condition \rf{lbc_inv} and derive the dispersion
relation for Riemann ellipsoids, it remains to obtain an explicit expression for the
pressure perturbation. This is achieved in the following section by deriving a
generalised Poincar\'e equation governing the normal modes of oscillation.

\subsection{Normal modes of oscillation}

We first introduce an inviscid hydrodynamic potential, simultaneously taking 
into account the effects of pressure and gravity,
\begin{equation}
\label{W}
\delta W^{(0)} = \delta\Phi^{(0)} - \delta p^{(0)} .
\end{equation}

Applying the divergence operator on \rf{leom1} and using \rf{leom2}, we 
obtain a general Poincar\'e equation governing the small oscillations of the 
rotating fluid,
\begin{equation}
\label{Poincare}
\mathcal{A}\{\delta\vu^{(0)}\} = \mathcal{P}\{\delta W^{(0)}\} .
\end{equation}
The Poincar\'e and advection operators are respectively given by
\begin{align}
\mathcal{P}\{\delta W^{(0)}\} &\eqdef \nab\scal\big( L^{-1} \nab \delta W^{(0)} \big) 
\eqdef \Tr \big[ (L^{-1})^T \Hess (\delta W^{(0)}) \big] , \\
\mathcal{A}\{\delta \vu^{(0)}\} &\eqdef \nab\scal\big[ L^{-1}
\big( \vu_0\scal\nab\delta\vu^{(0)} \big) \big] ,
\end{align}
where we have used the identity $\nab\scal(A\bm{q})=\Tr[A^T(\nab\otimes\bm{q})]$
that is valid for any square matrix $A$ and vector $\bm{q}$. Hess denotes the Hessian
matrix operator. The matrix $L$ is given by
\begin{equation}
L \eqdef
\begin{pmatrix}
-2\ui\omega & Q_1 - 2\Ro & 0 \\
Q_2 + 2\Ro & -2\ui\omega & 0 \\
0 & 0 & -2\ui\omega
\end{pmatrix} ,
\end{equation}
and is skew-Hermitian when $\omega\in\mathbb{R}$ and $Q_1=Q_2=0$.
In the absence of gravity, the spectrum of the Poincaré operator is known to be real
and bounded in the interval $[-\omega,\omega]$, with polynomial eigenfunctions
\citep{G65}. When dynamical instabilities occur (usually through a
Hamilton--Hopf bifurcation \citep{Kirillov}), the spectrum becomes
complex-valued and these properties no longer hold.

When $f = 0$, corresponding to a triaxial Jacobi ellipsoid, the problem reduces to 
$\mathcal{P}\{ W\}=0$, which is the classical Poincar\'e equation \citep{P85}. Its solutions 
form a one-parameter family of homogeneous polynomials of the form
$\alpha (x + \ui y)^n$, as established by \cite{C22}. Since the base flow $\vu_0$ 
depends linearly on the Cartesian coordinates, the advection operator $\mathcal{A}$ 
maps the space of Cartesian polynomials of degree $n-1$ onto itself. Accordingly, 
we seek solutions such that $\delta\vu^{(0)}\in\mathbb{P}^3_{n-1}$ and 
$\delta W^{(0)}\in\mathbb{P}^1_{n}$, where $\mathbb{P}^d_{n}$ denotes the 
$d$-dimensional space of polynomials of degree $n$.

An extension of Cartan’s polynomial family solving the generalised Poincar\'e equation
\rf{Poincare} is found to be
\begin{equation}
\label{Wpol}
\delta W^{(0)} (x,y,z) = - \alpha (x+\ui y)^n + \beta z^2 (x + \gamma y)^{n-2} ,
\end{equation}
where $(\alpha,\beta,\gamma)$ are real constants.
It can be shown that $\beta$ depends linearly on $\alpha$, implying that
\rf{Wpol} also constitutes a one-parameter family of solutions for Riemann ellipsoids.

Substituting this family into the general Poincaré equation \rf{Poincare}, the inviscid velocity 
perturbation $\delta\vu^{(0)} = (\delta u_x^{(0)}, \delta u_y^{(0)}, \delta u_z^{(0)})$
can be written in the form
\begin{equation}
\label{delta_u0}
\begin{gathered}
\delta u_x^{(0)} (x,y,z) = \sum_{k=0}^{n-1} \widehat{u}_k x^{n-1-k} y^{k} + \widehat{u}_n z^{2} , \quad
\delta u_y^{(0)} (x,y,z) = \sum_{k=0}^{n-1} \widehat{v}_k x^{n-1-k} y^{k} + \widehat{v}_n z^{2} , \\
\delta u_z^{(0)} (x,y,z) = z \sum_{k=0}^{n-2} \widehat{w}_k x^{n-2-k} y^{k} ,
\end{gathered}
\end{equation}
where the polynomial coefficients $(\widehat{u}_k,\widehat{v}_k,\widehat{w}_k)$ and the constants
$(\beta,\gamma)$ are determined by term-by-term identification with the Euler and continuity equations.
It can be shown that $(\beta,\gamma,\widehat{w}_k)$ depend linearly on the background vorticity $\zeta$.
Hence, in the limit $\zeta = 0$, the classical Cartan solutions parameterised by $\alpha$ are recovered.
This family of solutions, for arbitrary harmonic degree and for axisymmetric configurations 
(i.e. Maclaurin spheroids with $\Gamma = 1$) is given by \citep{C79},
\begin{equation}
\delta u_x^{(0)} = -\frac{\ui n\alpha(x+\ui y)^{n-1}}{2\Ro(\sigma+1)} , \quad
\delta u_y^{(0)} = \frac{n\alpha(x+\ui y)^{n-1}}{2\Ro(\sigma+1)} , \quad 
\delta u_z^{(0)} = 0 .
\end{equation}

For $n>2$, the velocity perturbations of Riemann ellipsoids are not harmonic.
This feature is a major obstacle to extend the classical viscous stability
theory developed for Maclaurin spheroids, as historically done by \cite{RS63}.
Since the potential $\delta W^{(0)}$ is also non-harmonic in the triaxial case,
the associated pressure perturbation
\begin{equation}
\delta p^{(0)} \eqdef \delta \Phi^{(0)} - \delta W^{(0)} = \delta \Phi^{(0)} + \alpha (x+\ui y)^n - \beta z^2 (x + \gamma y)^{n-2} ,
\end{equation}
is likewise non-harmonic for $n>2$. Here, $\delta\Phi^{(0)}$ is expressed in terms of the velocity field
\rf{delta_u0} via relations \rf{delta_phi0} and \rf{cte_xi_u}.

\subsection{Dispersion relation and stability results}

To evaluate the hydrodynamic potential at the boundary of the ellipsoid,
we expand expression \rf{Wpol} in terms of surface harmonics,
\begin{equation}
\label{delta_W0}
\big[ \delta W^{(0)} \big]_{\partial\mathcal{D}_0} = \widehat{W}_{nm} \mathbb{S}_n^m \ue^{-2\ui t\sigma\Ro} ,
\end{equation}
with expansion coefficients
\begin{equation}
\label{cte_W0}
\widehat{W}_{nm} = \frac{\ue^{2\ui t\sigma\Ro}}{\varpi_n^m} \int
\left[ - \alpha (x+\ui y)^n + \beta z^2 (x + \gamma y)^{n-2} \right]_{\partial\mathcal{D}_0} \mathbb{S}_n^m
\ud S .
\end{equation}

Substituting \rf{xi}, \rf{delta_phi0}, and \rf{delta_W0} into the inviscid boundary
condition \rf{lbc_inv} yields the dispersion relation for inviscid Riemann ellipsoids,
\begin{equation}
\label{disp_rel_inv}
\pazocal{D}_n^m(\sigma) = 2 \theta_n^m \widehat{\xi}_{nm} + \widehat{W}_{nm} = 0 ,
\end{equation}
where
\begin{equation}
\theta_n^m \eqdef P_0 - 2 \pazocal{W}(1)^{-1} E_n^m(1) F_n^m(1).
\end{equation}

Using expressions \rf{cte_xi_u} and \rf{cte_W0}, the dispersion relation can be written
explicitly as
\begin{align}
\label{disp_rel_inv_2}
\pazocal{D}^{(0)}(\sigma) = &\frac{\ui\theta_n^m}{\sigma\Ro} \int
\left[ \frac{\partial x}{\partial\lambda} \delta u_x^{(0)} + \frac{\partial y}{\partial\lambda} \delta u_y^{(0)}
+ \frac{\partial z}{\partial\lambda} \delta u_z^{(0)} \right]_{\partial\mathcal{D}_0} \mathbb{S}_n^m \ud S \\
&- \int\left[ \alpha (x+\ui y)^n - \beta z^2 (x + \gamma y)^{n-2} 
\right]_{\partial\mathcal{D}_0} \mathbb{S}_n^m \ud S = 0 . \nonumber
\end{align}

Once the degree and order of the ellipsoidal harmonic are specified, the associated 
Lam\'e functions are computed following the procedure described in Appendix~\ref{app3}.
The linear stability of the system then follows directly from solving
\rf{disp_rel_inv_2} for the eigenfrequency $\sigma$.

Finally, setting $f=0$ (or equivalently $\zeta = 0$) in \rf{disp_rel_inv_2} 
yields the classical dispersion relation for Jacobi ellispoids \citep{C62}.
In addition, setting $\Gamma = 1$ yields the characteristic equation for the stability
of Maclaurin spheroids \citep{EFE,C79}
\begin{equation}
\pazocal{D}^{(0)}_{\textrm{Mc}}(\sigma) = \Ro^2\sigma(\sigma+1) - \theta_n^m = 0 .
\end{equation}

\begin{figure}[t!]
    \includegraphics[width=.33\textwidth]{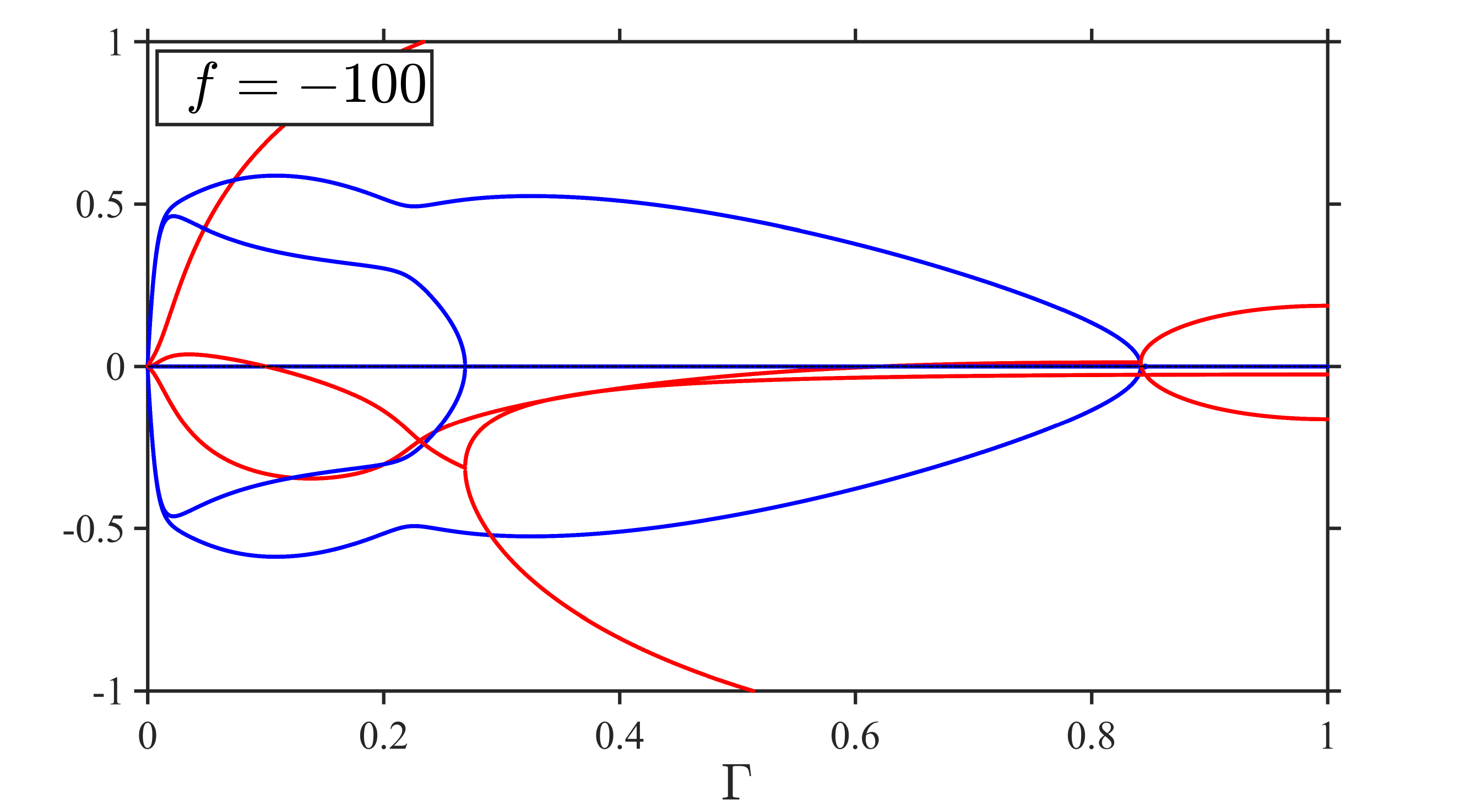}~
    \includegraphics[width=.33\textwidth]{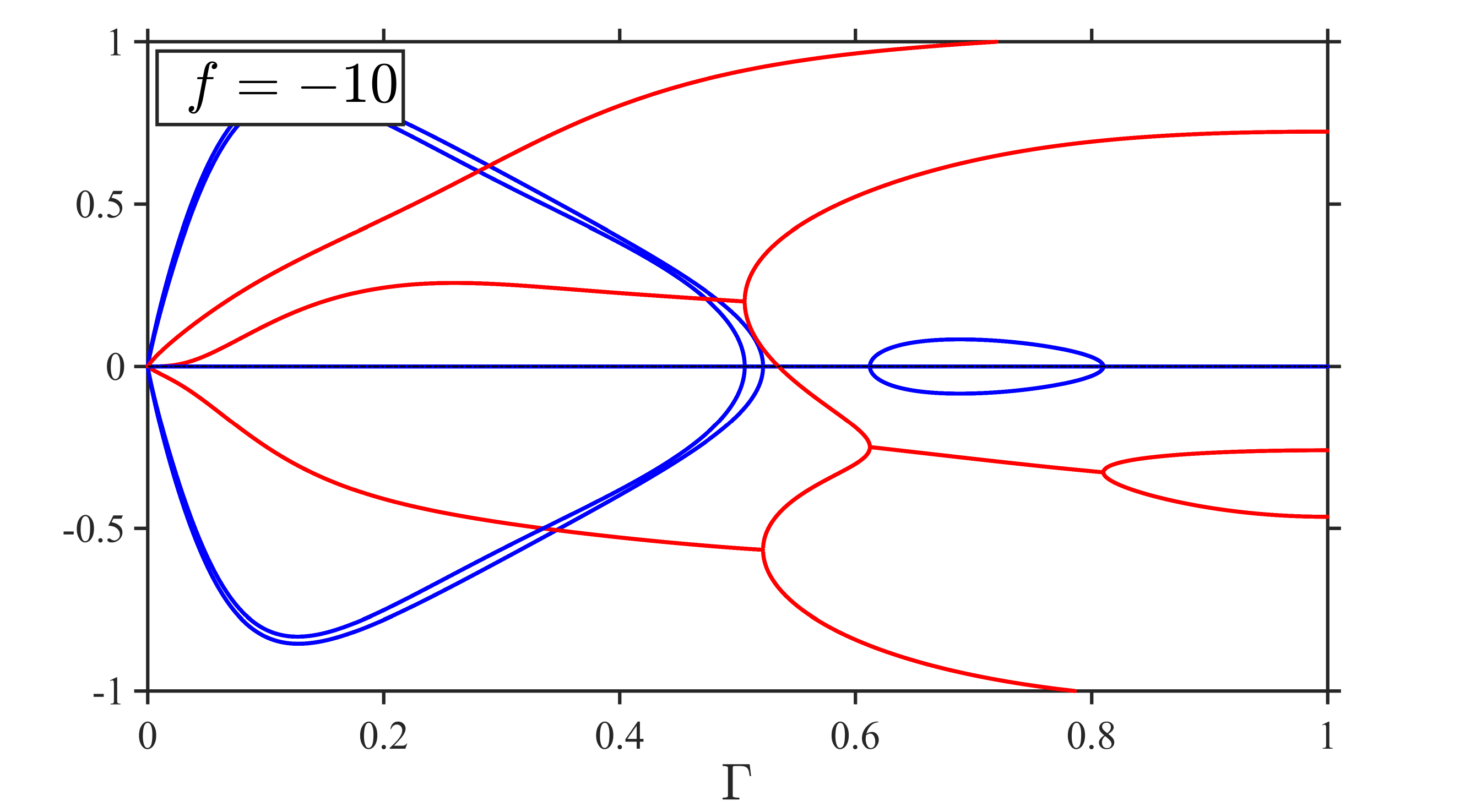}~
    \includegraphics[width=.33\textwidth]{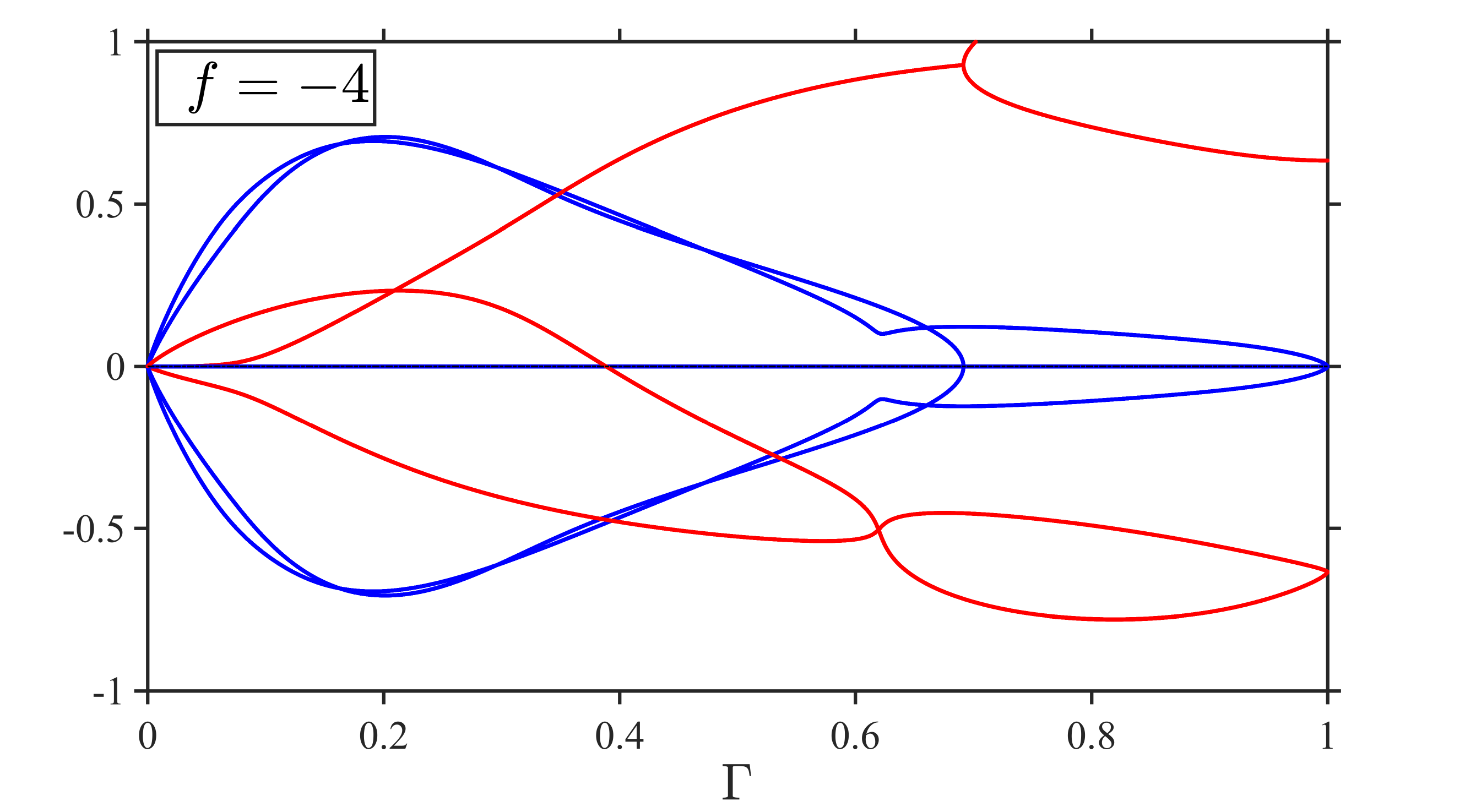}\\
    \includegraphics[width=.33\textwidth]{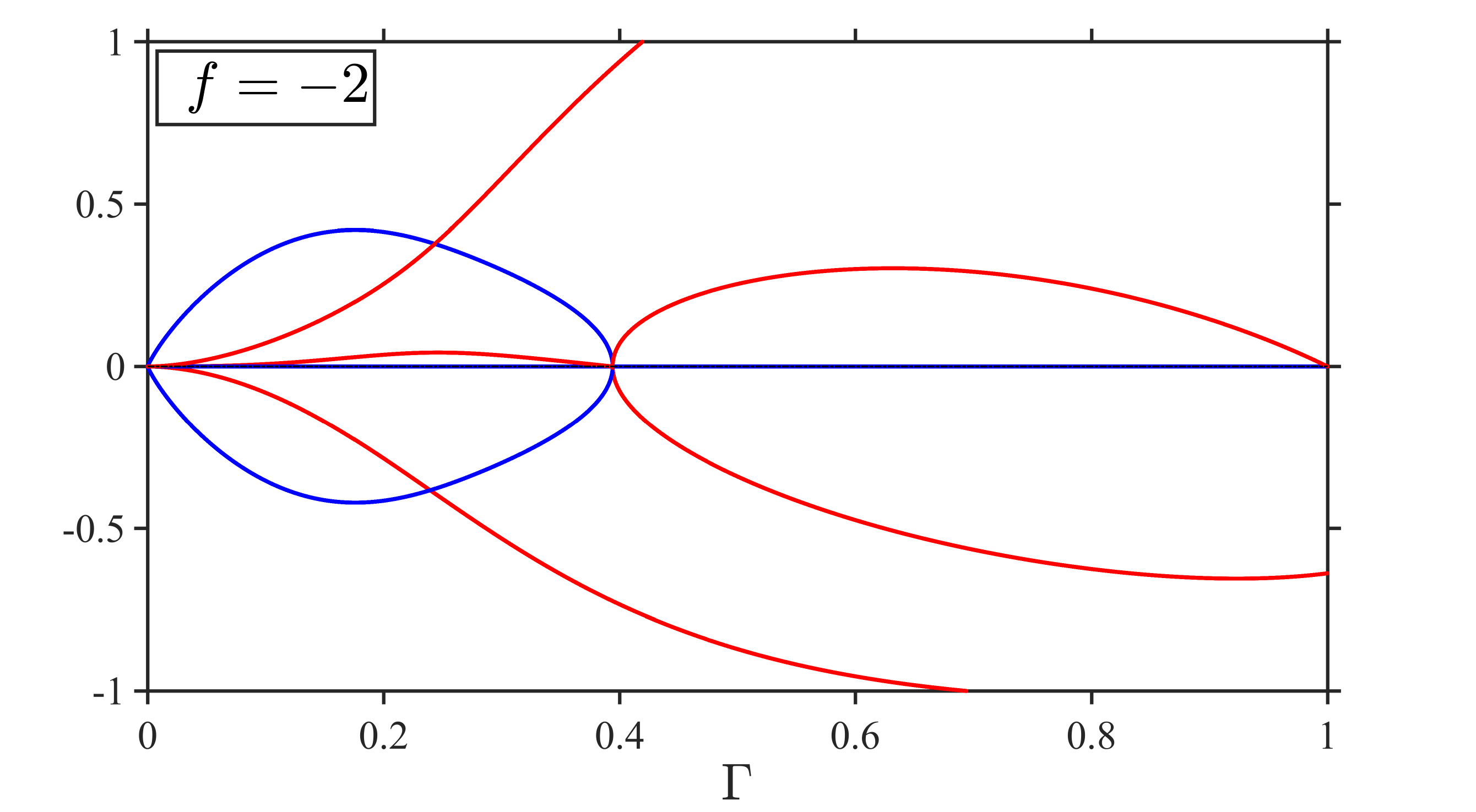}~
    \includegraphics[width=.33\textwidth]{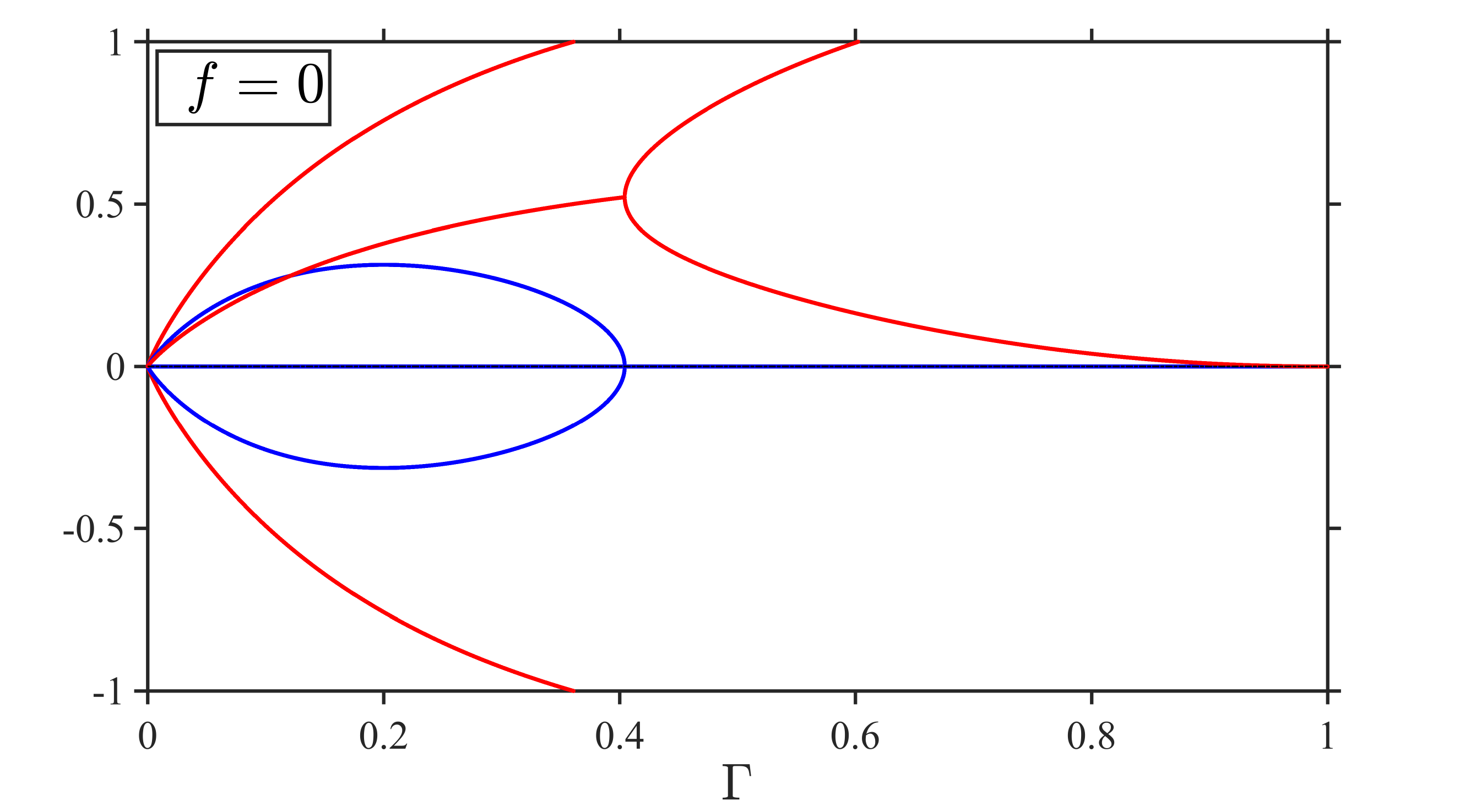}~
    \includegraphics[width=.33\textwidth]{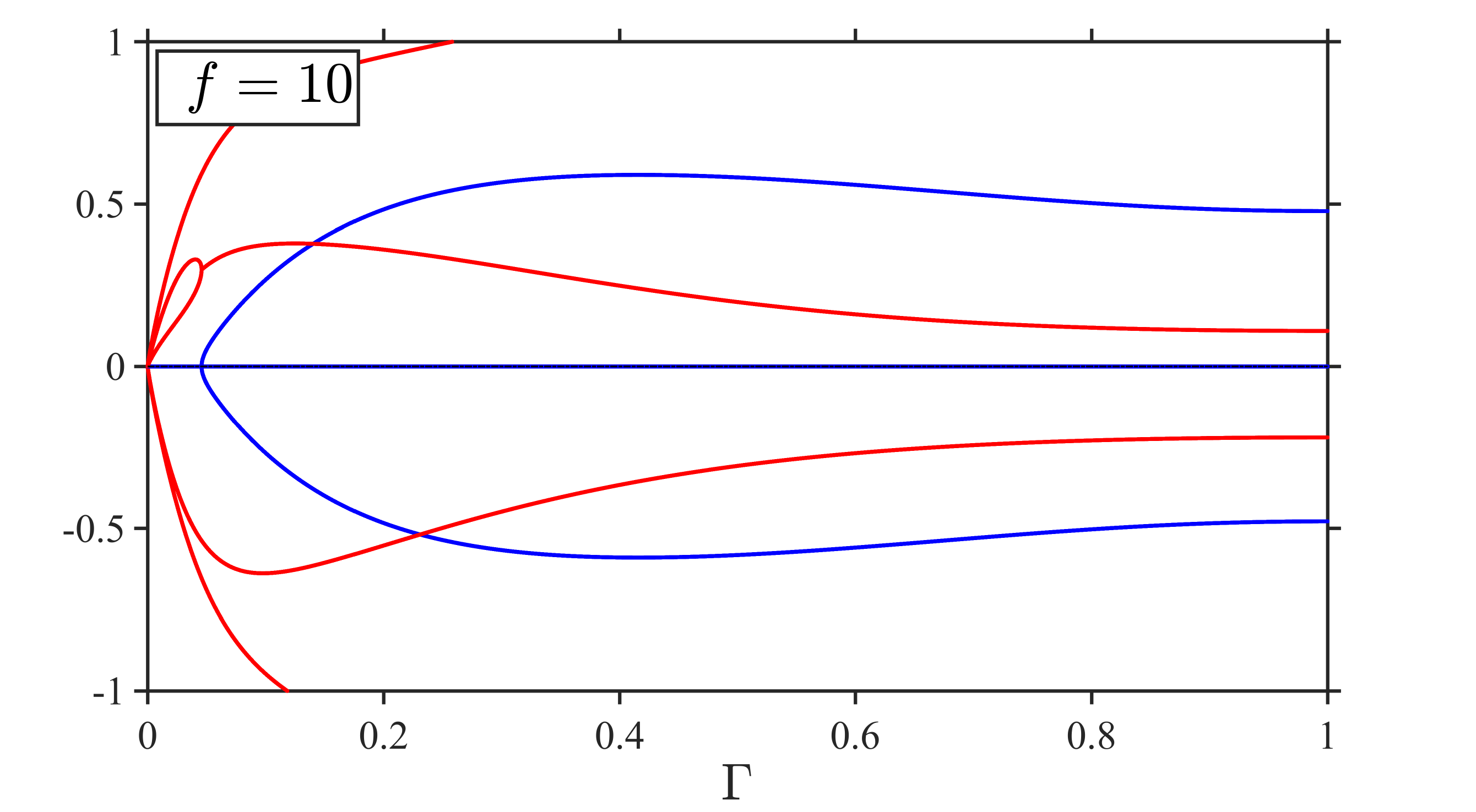}
    \caption{Dispersion curves of the $m=n=2$ inviscid Riemann ellipsoid
    over the planar aspect ratio $\Gamma$. Red (blue) lines denote the
    frequencies (growth rates) of the system, computed at different values of
    $f$ according to the legend.
    \label{Fig_inv}}
\end{figure}

We present in Figure~\ref{Fig_inv} a representative set of dispersion curves 
for the case $m=n=2$. This mode corresponds to the classical \textit{bar mode}, 
which is known to be the most unstable configuration for Maclaurin spheroids \citep{EFE}. 
In the limit $f=0$, we recover the stability diagram of the Jacobi ellipsoid, 
including the bifurcation point located at $\crit{\Gamma} \approx 0.4635$ and 
$\crit{\Xi} \approx 0.3632$. The value of $\cos^{-1}(\crit{\Xi})$ at this 
critical point is tabulated in \cite{C62}, and our result agrees with this
value to within less than $1^\circ$.

As $f$ departs from zero, similar bifurcations persist for $f>0$; however, 
the ellipsoid remains linearly unstable beyond the critical threshold. 
For $f<0$, additional instabilities arise through the crossings of distinct 
frequency branches. These unstable modes are attributed to the elliptic 
instability induced by the non-axisymmetric deformation of the internal 
streamlines, as discussed in \citep{K02}.

\section{Viscous Riemann ellipsoids} \label{sec_vis}

In this section, we investigate the effect of weak viscosity by assuming $\Ek \ll 1$.
We restrict our attention to the case $n = 2$, for which the inviscid velocity
field \rf{delta_u0} remains harmonic (see Section~\ref{sec_inv}).
For higher harmonic degrees ($n>2$), the inviscid solutions no longer satisfy
the viscous equations of motion, and $O(\Ek)$ corrections to the velocity field
are required for compatibility. By contrast, when $n=2$, viscous effects enter 
the problem solely through the boundary conditions.

In the present framework, viscous corrections arise from the requirement that
the stress-free boundary conditions be satisfied to first order in $\Ek$. This is 
achieved by introducing a thin viscous boundary-layer at the inner edge of 
the Riemann ellipsoid and solving the associated Prandtl boundary-layer equations.
This approach allows us to compute the leading-order viscous corrections to the normal 
stress condition and to derive an analytic dispersion relation, accounting for viscous effects.

\subsection{Surface stress conditions}

We start by enforcing the normal stress boundary condition \rf{lbc}, 
which is decomposed as
\begin{equation}
\label{vis_BC}
\big[ \delta p^{(0)} + \bm{\xi}^{(0)}\scal\nab p_0 \big]_{\partial\mathcal{D}_0} 
+ \big[ \delta p^{(\nu)} + \bm{\xi}^{(\nu)}\scal\nab p_0 - 2\Ro\Ek\varepsilon_{\lambda\lambda}^{(0)} \big] = 0 ,
\end{equation}
where the first term in brackets corresponds to the inviscid contribution, which
is determined by the dispersion relation obtained in Section~\ref{sec_inv}.
The second term in brackets is anticipated to be of order $O(\Ek)$.
As a consequence, the viscous contribution arising from the Cauchy stress
tensor involves only the inviscid normal velocity field. 

Expanding \rf{vis_BC} in surface harmonics and integrating over the equilibrium
boundary yields
\begin{equation}
\label{vis_BC_ell}
\int \big[ \delta p^{(0)} + \bm{\xi}^{(0)}\scal\nab p_0 
- 2\Ro\Ek \varepsilon_{\lambda\lambda}^{(0)} \big]_{\partial\mathcal{D}_0} \mathbb{S}_n^m \ud S
+ \int \big[ \delta p^{(\nu)} + \bm{\xi}^{(\nu)}\scal\nab p_0
\big]_{\partial\mathcal{D}_0} \mathbb{S}_n^m \ud S = 0 ,
\end{equation}
where the first integral is evaluated using the inviscid analysis developed in
Section~\ref{sec_inv}.

Using the viscous counterpart of \rf{xi_def}, together with the definition of
the perturbation to the gravitational potential given in \rf{delta_phi0}, the
second integral may be rewritten as
\begin{equation}
\label{Iint}
\mathfrak{I} = \int \Big[ \delta p^{(\nu)} - \delta \Phi^{(\nu)} - \frac{\ui\theta_n^m}{\sigma\Ro} 
h_\lambda \delta u_\lambda^{(\nu)} \Big]_{\partial\mathcal{D}_0} \mathbb{S}_n^m \ud S .
\end{equation}

The objective of the present section is to evaluate this integral.
To this end, we introduce a viscous boundary-layer formulation near the inner
surface of the ellipsoid, following the approach
previously employed for Maclaurin spheroids by \cite{RS63} and \cite{C79}.

\subsection{Boundary-layer theory} \label{subsec_BL}

We assume that advection in the direction normal to the boundary is negligible
at leading-order in viscosity.
In addition, we require the viscous correction terms to vanish at the inner
surface of the ellipsoid and to asymptotically match the inviscid solution away
from the boundary layer.
Denoting the characteristic thickness of the layer by
$d_\nu = O(\Ek^{1/2})$, classical boundary-layer scaling arguments
\citep{LandauLifshitz} follow from enforcing consistency between the equations of
motion and the boundary conditions at higher order in $\Ek$.
This yields
\begin{gather*}
\frac{\partial}{\partial\lambda} = O(\Ek^{-1/2}), \quad
\frac{\partial}{\partial\mu} = O(1), \quad
\frac{\partial}{\partial\varphi} = O(1),
\end{gather*}
together with
\begin{gather*}
\delta u_\lambda^{(\nu)} = O(\Ek), \;
\delta u_\mu^{(\nu)} = O(\Ek^{1/2}), \;
\delta u_\varphi^{(\nu)} = O(\Ek^{1/2}), \;
\delta p^{(\nu)} = O(\Ek), \;
\delta \Phi^{(\nu)} = O(\Ek), \\
u_{0\lambda} = O(\Ek^{1/2}), \quad
u_{0\mu} = O(1), \quad
u_{0\varphi} = O(1),
\end{gather*}
where the scaling of the background velocity field
$\vu_0 \eqdef (u_{0\lambda}, u_{0\mu}, u_{0\varphi})$
follows from the incompressibility constraint.

We now expand the incompressible Navier--Stokes equations \rf{leom1} together
with the continuity equation \rf{leom2} in powers of $\Ek$ and retain the 
leading-order contributions in each expression.
Projecting the equations onto the normal direction and enforcing the 
solenoidal condition yields, respectively,
\begin{align}
h_\mu h_\varphi \frac{\partial}{\partial\lambda} \Big[ \delta p^{(\nu)} - \delta \Phi^{(\nu)} \Big] &= 
2\Ro h_\lambda \Big[ \frac{\partial z}{\partial \varphi} h_\mu \delta u_\mu^{(\nu)} - 
\frac{\partial z}{\partial \mu} h_\varphi \delta u_\varphi^{(\nu)} \Big] ,
\label{vleomn} \\
h_\mu h_\varphi \frac{\partial}{\partial\lambda} \delta u_\lambda^{(\nu)} &= 
- \Big[ \frac{\partial}{\partial \mu} \big( h_\lambda h_\varphi \delta u_\mu^{(\nu)} \big) + 
\frac{\partial}{\partial \varphi} \big( h_\lambda h_\mu \delta u_\varphi^{(\nu)} \big) \Big] .
\label{vleomdiv}
\end{align}

Combining \rf{vleomn} and \rf{vleomdiv} so as to recover the integrand of
$\mathfrak{I}$, we obtain
\begin{gather}
\label{diff_I}
h_\mu h_\varphi \frac{\partial}{\partial\lambda} \Big[ \delta p^{(\nu)} - \delta \Phi^{(\nu)} 
- \frac{\ui\theta_n^m}{\sigma\Ro} h_\lambda\delta u_\lambda^{(\nu)} \Big] =
2\Ro h_\lambda \Big[ \frac{\partial z}{\partial \varphi} h_\mu \delta u_\mu^{(\nu)} - 
\frac{\partial z}{\partial \mu} h_\varphi \delta u_\varphi^{(\nu)} \Big] \\
+ \frac{\ui\theta_n^m}{\sigma\Ro} h_\lambda \Big[ \frac{\partial}{\partial \mu} 
\big( h_\lambda h_\varphi \delta u_\mu^{(\nu)} \big) + \frac{\partial}{\partial \varphi} 
\big( h_\lambda h_\mu \delta u_\varphi^{(\nu)} \big) \Big] . \nonumber
\end{gather}
It is important to notice that every geometric terms (e.g. the scale factors) are
evaluated at the boundary and thus, are independent of the normal coordinate $\lambda$.

We now expand the viscous corrections to the tangential velocity components in
terms of surface harmonics, allowing for an arbitrary dependence on the normal
coordinate. This yields
\begin{equation}
\label{u_tangent}
\delta u_\mu^{(\nu)} = \Ek^{1/2} \widehat{u}_\mu (\lambda) \mathbb{S}_n^m (\mu,\varphi) , 
\quad
\delta u_\varphi^{(\nu)} = \Ek^{1/2} \widehat{u}_\varphi (\lambda) \mathbb{S}_n^m (\mu,\varphi) .
\end{equation}
Substituting this ansatz into \rf{diff_I}, multiplying by the appropriate metric
factors, integrating over the surface element \rf{surf_elem}, and invoking
Leibniz’s rule (since all $\lambda$-dependent quantities other than the velocity
corrections are constant), we obtain
\begin{equation}
\label{dIdlambda}
\frac{\ud\mathfrak{I}}{\ud\lambda} = \Ek^{1/2} \big[ \widehat{u}_\mu (\lambda) \mathfrak{I}_\mu +
\widehat{u}_\varphi (\lambda) \mathfrak{I}_\varphi \big] ,
\end{equation}
where the coefficients are explicitly given by
\begin{align}
\mathfrak{I}_\mu \eqdef &\;2\Ro \int \frac{h_\lambda}{h_\varphi} \frac{\partial z}{\partial\varphi} 
\big[ \mathbb{S}_n^m \big]^2 \ud S
+ \frac{\ui\theta_n^m}{\sigma\Ro} \int \frac{h_\lambda}{h_\mu h_\varphi} \frac{\partial}{\partial\mu} 
\big( h_\lambda h_\varphi \mathbb{S}_n^m \big) \mathbb{S}_n^m \ud S ,
\\
\mathfrak{I}_\varphi \eqdef &\;2\Ro \int \frac{h_\lambda}{h_\mu} \frac{\partial z}{\partial\mu} 
\big[ \mathbb{S}_n^m \big]^2 \ud S
+ \frac{\ui\theta_n^m}{\sigma\Ro} \int \frac{h_\lambda}{h_\mu h_\varphi} \frac{\partial}{\partial\varphi} 
\big( h_\lambda h_\mu \mathbb{S}_n^m \big) \mathbb{S}_n^m \ud S .
\end{align}
As anticipated, $\mathfrak{I}$ is of order $O(\Ek)$.

We now seek explicit expressions for the expansion coefficients
$(\widehat{u}_\mu,\widehat{u}_\varphi)$.
To this end, we project the Navier--Stokes equations onto the tangent plane and
retain the leading-order viscous contributions.
This procedure yields a coupled system of Prandtl-type equations,
\begin{subequations}
\label{Prandtl}
\begin{align}
\bigg[ -2\ui\sigma\Ro + \vu_0\scal\nab - \frac{\Ro\Ek}{h_\lambda^2} \frac{\partial^2}{\partial \lambda^2}
+ \frac{1}{h_\mu} \frac{\partial u_{0\mu}}{\partial\mu} \bigg] \delta u_\mu^{(\nu)} 
+ \bigg[ \frac{1}{h_\varphi} \frac{\partial u_{0\mu}}{\partial\varphi} 
- \frac{2\Ro}{h_\lambda} \frac{\partial z}{\partial \lambda} \bigg] \delta u_\varphi^{(\nu)} 
&= 0 , \\
\bigg[ -2\ui\sigma\Ro + \vu_0\scal\nab - \frac{\Ro\Ek}{h_\lambda^2} \frac{\partial^2}{\partial \lambda^2}
+ \frac{1}{h_\varphi} \frac{\partial u_{0\varphi}}{\partial\varphi} \bigg] \delta u_\varphi^{(\nu)} 
+ \bigg[ \frac{1}{h_\mu} \frac{\partial u_{0\varphi}}{\partial\mu} 
+ \frac{2\Ro}{h_\lambda} \frac{\partial z}{\partial \lambda} \bigg] \delta u_\mu^{(\nu)} 
&= 0 .
\end{align}
\end{subequations}
In the absence of prevalent motion, the system \rf{Prandtl} reduces to a pair of
coupled non-homogeneous heat equations whose solutions are described analytically; 
see \cite{RS63} or \cite{C79}.

\subsection{Analytic solutions and secular instability}

We require the viscous correction fields to decay away from the boundary so that
the velocity field asymptotically matches the inviscid solution in the bulk of
the ellipsoid.
This condition is expressed as
\begin{equation}
\label{odes_bc1}
\lim_{\eta\to-\infty} \big[ \widehat{u}_\mu \big] =
\lim_{\eta\to-\infty} \big[ \widehat{u}_\varphi \big] = 0 ,
\end{equation}
where we introduced a stretched normal coordinate
\begin{equation}
\eta \eqdef d_\nu^{-1} (\lambda-1) = \frac{\lambda-1}{\Ek^{1/2}} \in (-\infty,0] ,
\end{equation}
which captures variations across the boundary layer.

Substituting \rf{u_tangent} into the system \rf{Prandtl} and
integrating over the tangential coordinates yields a coupled system of linear
ordinary differential equations with constant coefficients,
\begin{subequations}
\label{odes}
\begin{align}
a_{0} \frac{\ud^2}{\ud\eta^2} \widehat{u}_\mu (\eta) + a_{1} \frac{\ud}{\ud\eta} \widehat{u}_\mu (\eta) +
a_{2} \widehat{u}_\mu (\eta) + a_{3} \widehat{u}_\varphi (\eta) &= 0 , \\
b_{0} \frac{\ud^2}{\ud\eta^2} \widehat{u}_\varphi (\eta) + b_{1} \frac{\ud}{\ud\eta} \widehat{u}_\varphi (\eta) +
b_{2} \widehat{u}_\varphi (\eta) + b_{3} \widehat{u}_\mu (\eta) &= 0 ,
\end{align}
\end{subequations}
where the boundary-layer scaling introduced in Section~\ref{subsec_BL} has been
used to isolate the viscous contribution from the global dynamics.

Similarly, integrating the tangential stress boundary conditions \rf{lbc} yields
\begin{equation}
\label{odes_bc2}
\lim_{\eta=0} \Big[ \varpi_n^m \frac{\ud}{\ud\eta} \widehat{u}_\mu + \tau_\mu \Big] = 
\lim_{\eta=0} \Big[ \varpi_n^m \frac{\ud}{\ud\eta} \widehat{u}_\varphi + \tau_\varphi \Big] = 0 .
\end{equation}
Explicit expressions for the coefficients
$(a_j, b_j, \tau_\mu, \tau_\varphi)$ are provided in Appendix~\ref{app4}.

Introducing the state vector $\bm{\xi}\eqdef(\widehat{u}_\mu,\widehat{u}_\varphi)$,
the system \rf{odes} can be written compactly as
\begin{equation}
\label{odes_lin}
\begin{pmatrix}
a_0 & 0 \\
0 & b_0
\end{pmatrix} \frac{\ud^2\bm{\xi}}{\ud\eta^2} +
\begin{pmatrix}
a_1 & 0 \\
0 & b_1
\end{pmatrix} \frac{\ud\bm{\xi}}{\ud\eta} +
\begin{pmatrix}
a_2 & a_3 \\
b_3 & b_2
\end{pmatrix} \bm{\xi} \eqdef
\bm{A}_0 \frac{\ud^2\bm{\xi}}{\ud\eta^2} + \bm{A}_1
\frac{\ud\bm{\xi}}{\ud\eta} + \bm{A}_2 \bm{\xi} = \bm{0} ,
\end{equation}
supplemented with boundary conditions \rf{odes_bc1} and \rf{odes_bc2}.
This linear system admits exponentially decaying solutions of the form
$\bm{\xi} \sim \exp(\kappa \eta)$.
The corresponding characteristic equation is
\begin{equation}
\label{char_pol}
P(\kappa) = \det\left(\kappa^2 \bm{A}_0 - \kappa \bm{A}_1 + \bm{A}_0\right) = 0 ,
\end{equation}
which is a quartic polynomial in $\kappa$ whose roots 
generally possess both positive and negative real parts.

Since physical admissibility requires decay away from the boundary,
only roots with positive real parts are retained.
Denoting these roots by $\kappa_1$ and $\kappa_2$,
the tangential velocity corrections may be written as
\begin{equation}
\label{odes_sol}
\widehat{u}_\mu (\eta) = A_1 \ue^{\kappa_1\eta} + A_2 \ue^{\kappa_2\eta} , 
\quad
\widehat{u}_\varphi (\eta) = B_1 \ue^{\kappa_1\eta} + B_2 \ue^{\kappa_2\eta} .
\end{equation}

The constants $K_j$ are obtained by enforcing the boundary conditions
\rf{odes_bc1}--\rf{odes_bc2} and substituting \rf{odes_sol} into
\rf{odes_lin}, which yields the linear system
\begin{equation}
\label{syst_Kj}
\begin{pmatrix}
A_1 \\ A_2 \\ B_1 \\ B_2
\end{pmatrix} =
\begin{pmatrix}
\mathcal{P}_a(\kappa_1) & \mathcal{P}_a(\kappa_2) & a_3 & a_3 \\
b_3 & b_3 & \mathcal{P}_b(\kappa_1) & \mathcal{P}_b(\kappa_2) \\
-\varpi_n^m \kappa_1 & -\varpi_n^m \kappa_2 & 0 & 0 \\
0 & 0 & -\varpi_n^m \kappa_1 & -\varpi_n^m \kappa_2
\end{pmatrix}^{-1}
\begin{pmatrix}
0 \\
0 \\
\tau_\mu \\
\tau_\varphi
\end{pmatrix} ,
\end{equation}
where $\mathcal{P}_a(\kappa)\eqdef a_0 \kappa^2 + a_1 \kappa + a_2$ and 
$\mathcal{P}_b(\kappa)\eqdef b_0 \kappa^2 + b_1 \kappa + b_2$.

The integral $\mathfrak{I}$ entering the normal stress condition is
then obtained by integrating \rf{dIdlambda} over the stretched coordinate $\eta$
and substituting the expressions for $\kappa_j$ and $K_j$.
This yields
\begin{align*}
\mathfrak{I} = \Ek \Big[ \mathfrak{I}_\mu \int_{-\infty}^{0} \widehat{u}_\mu \ud\eta  
+ \mathfrak{I}_\varphi \int_{-\infty}^{0} \widehat{u}_\varphi \ud\eta \Big]
= \frac{\Ek}{\kappa_1\kappa_2} \Big[ \mathfrak{I}_\mu \big( \kappa_2 A_1 + \kappa_1 A_2 \big)
+ \mathfrak{I}_\varphi \big( \kappa_2 B_1 + \kappa_1 B_2 \big) \Big].
\end{align*}

The dispersion relation governing the stability of viscous Riemann ellipsoids
is given by
\begin{equation}
\label{vis_DR}
\pazocal{D}(\sigma) = \pazocal{D}^{(0)}(\sigma) + \Ek \pazocal{D}^{(\nu)}(\sigma) = 0 ,
\end{equation}
where $\pazocal{D}^{(0)}$ is obtained from \rf{disp_rel_inv_2} and 
\begin{equation}
\pazocal{D}^{(\nu)}(\sigma) = -2\Ro \int 
\big[ \varepsilon_{\lambda\lambda}^{(0)} \big]_{\pazocal{D}_0} \mathbb{S}_n^m \ud S
+ \mathfrak{I}_\mu \Big( \frac{A_1}{\kappa_1} + \frac{A_2}{\kappa_2} \Big)
+ \mathfrak{I}_\varphi \Big( \frac{B_1}{\kappa_1} + \frac{B_2}{\kappa_2} \Big) .
\end{equation}

We summarise the procedure for performing the linear stability analysis as follows:
\begin{itemize}
\item[1.] Fix the degree and order $(n,m)$ of the perturbation and compute the
corresponding ellipsoidal harmonic $\mathbb{S}_n^m$, as described in
Appendix~\ref{app3}.
\item[2.] Determine the polynomial coefficients of the inviscid velocity field
\rf{delta_u0} following the procedure outlined in
Section~\ref{sec_inv}. The case $n=2$ is already provided explicitly.
\item[3.] Compute the roots of the characteristic polynomial \rf{char_pol} with
positive real parts and determine the associated coefficients $K_j$ by solving
the linear system \rf{syst_Kj}.
\item[4.] Substitute the resulting expressions into the viscous dispersion relation
\rf{vis_DR} and solve for the complex frequency $\sigma$ using a suitable
root-finding algorithm. The stability of the system follows from the sign
of the growth rate.
\end{itemize}

As mentioned at the beginning of this section, the above procedure applies directly
only to the case $n = 2$, for which the inviscid velocity perturbations
\rf{delta_u0} satisfy the Navier--Stokes equations exactly.
For $n>2$, additional $O(\Ek)$ corrections to the bulk velocity field would be
required, since the inviscid solutions are no longer harmonic.
Although the present analysis is not extended to that case, the
methodology would remain formally unchanged.

\begin{figure}[t!]
    \includegraphics[width=.33\textwidth]{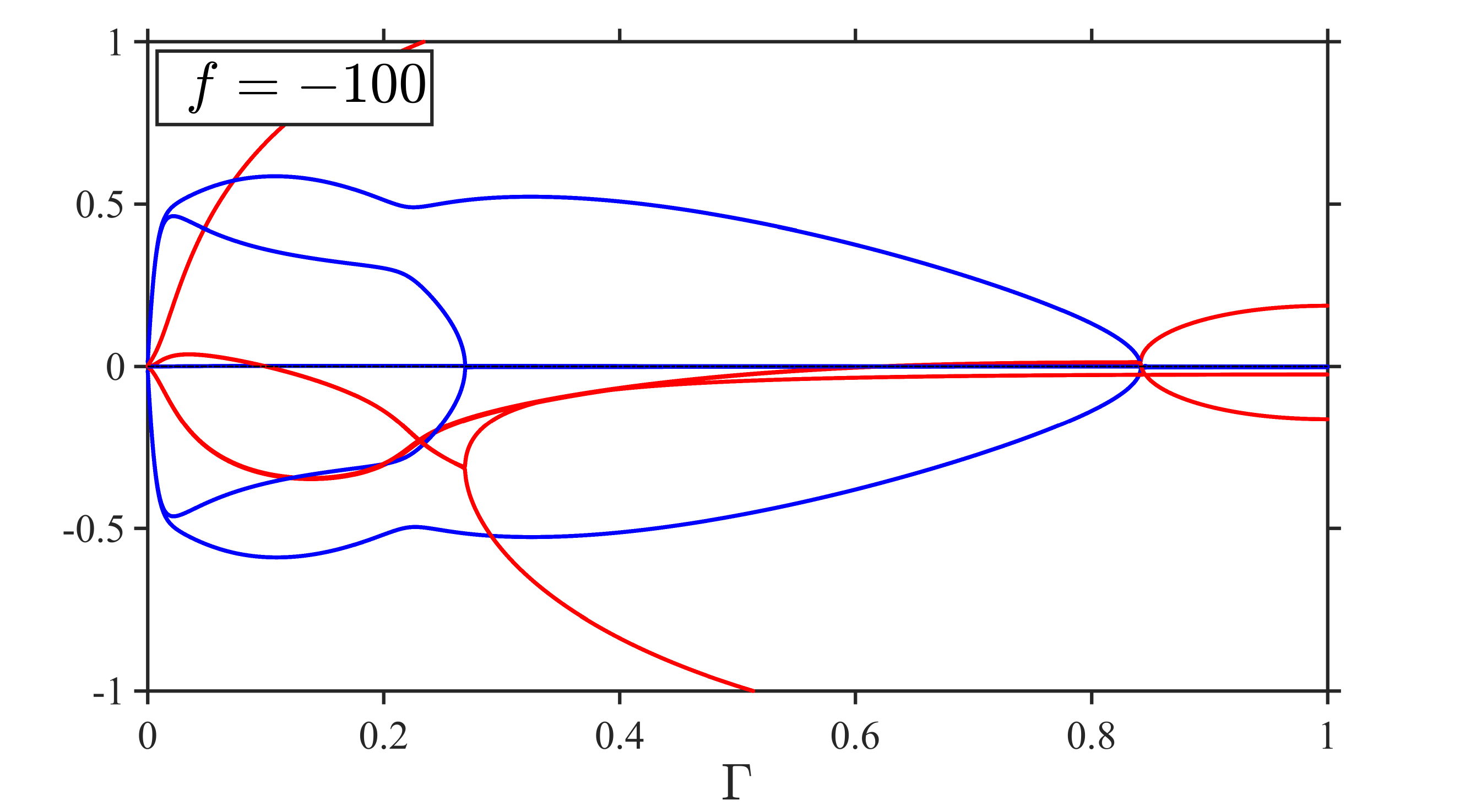}~
    \includegraphics[width=.33\textwidth]{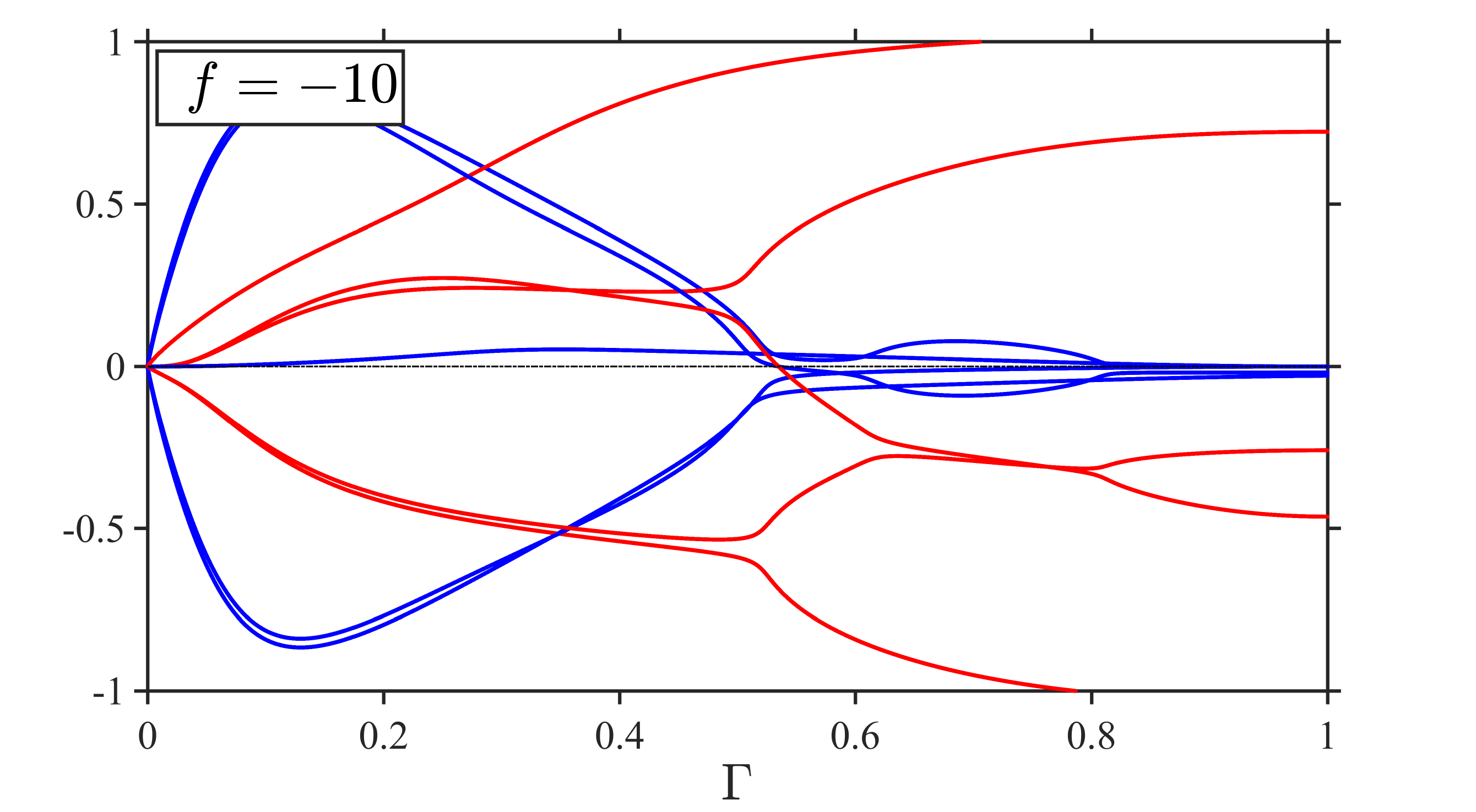}~
    \includegraphics[width=.33\textwidth]{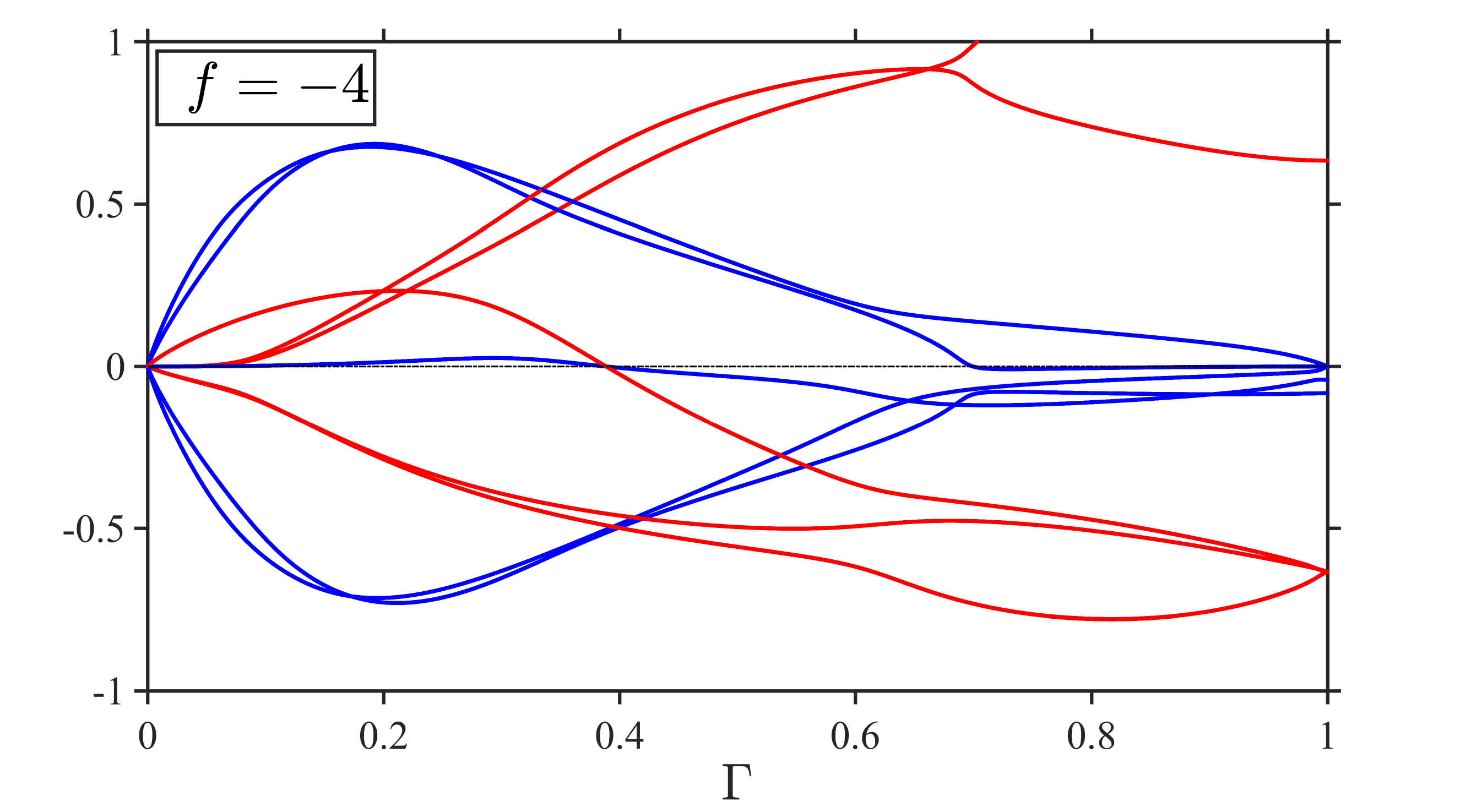}\\
    \includegraphics[width=.33\textwidth]{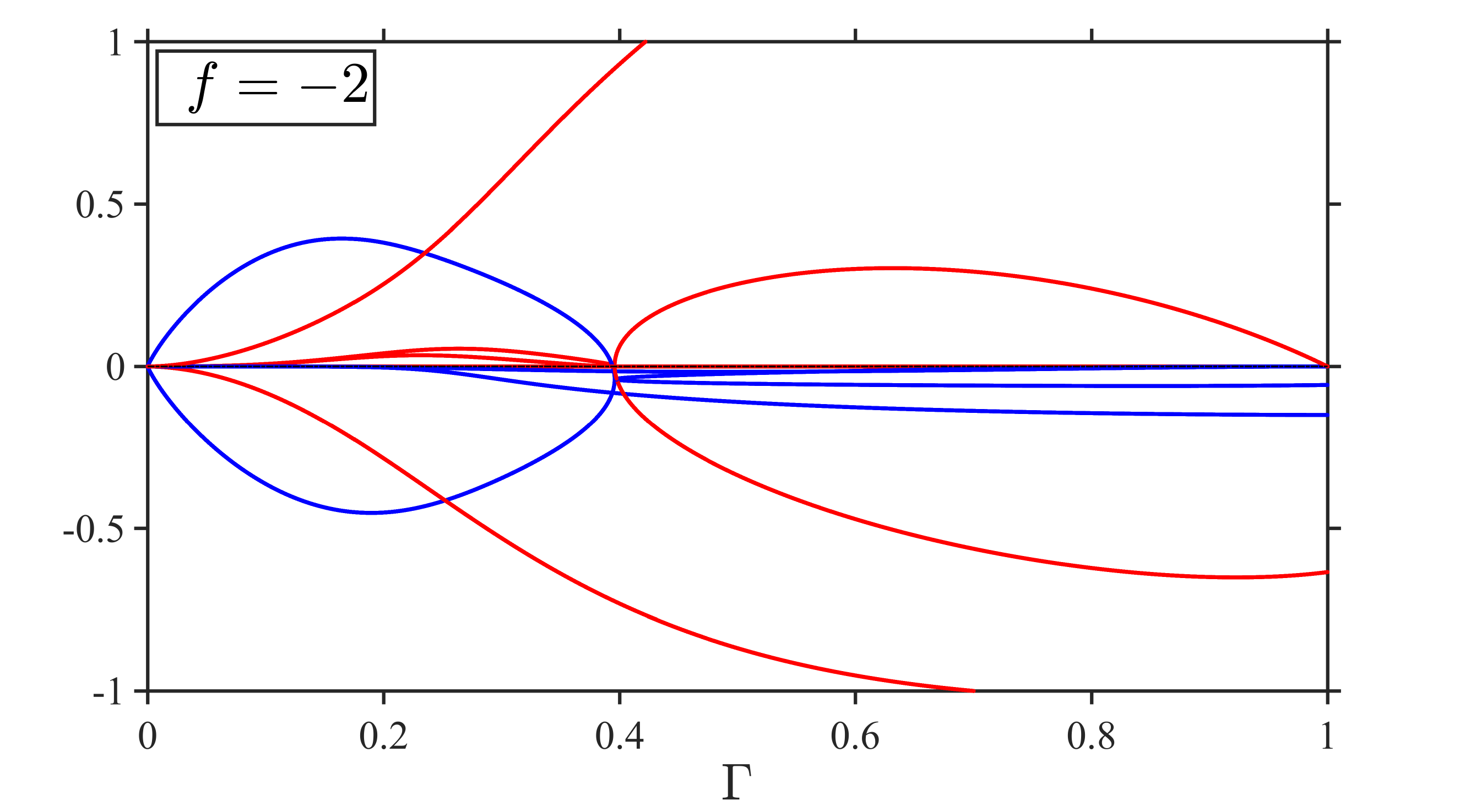}~
    \includegraphics[width=.33\textwidth]{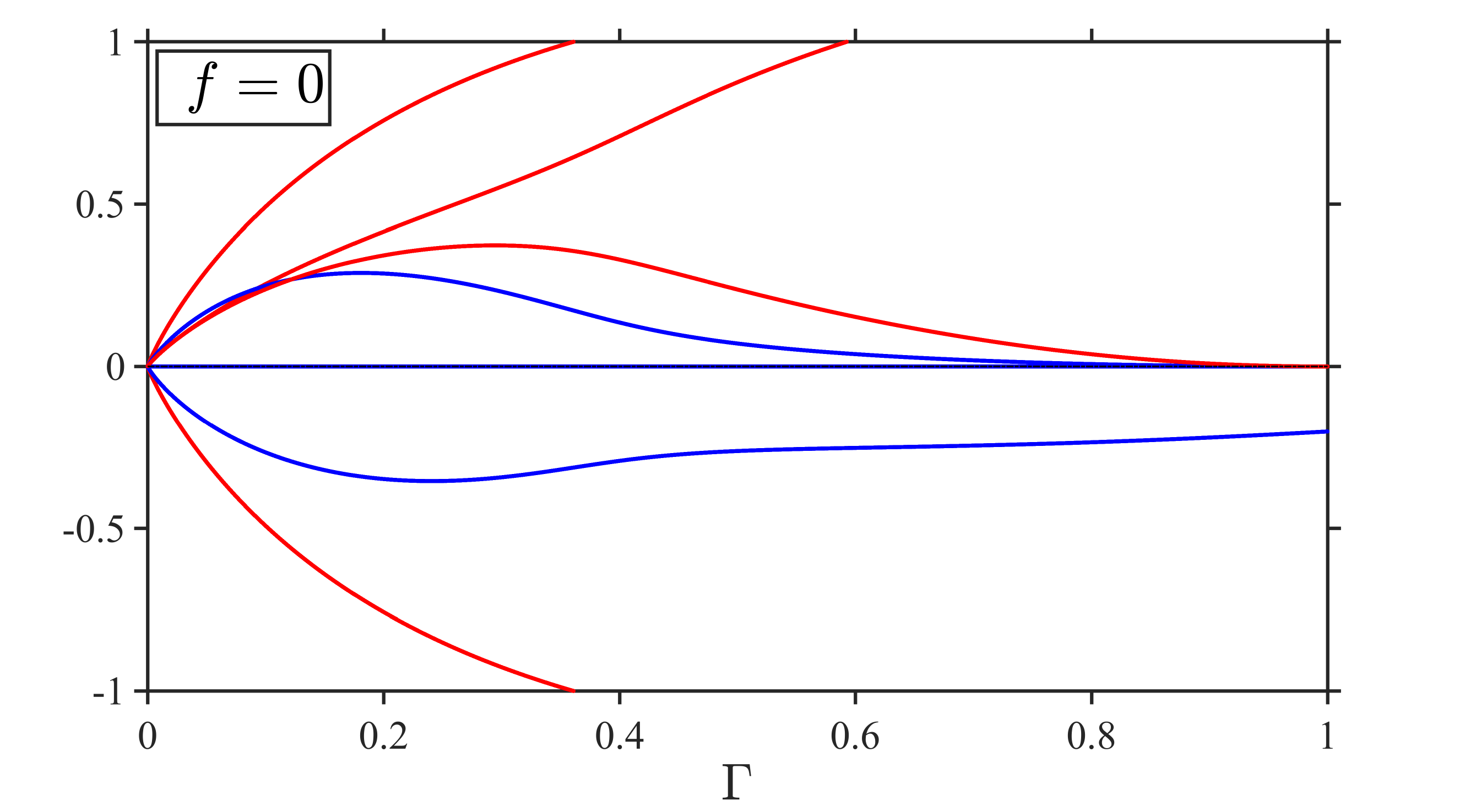}~
    \includegraphics[width=.33\textwidth]{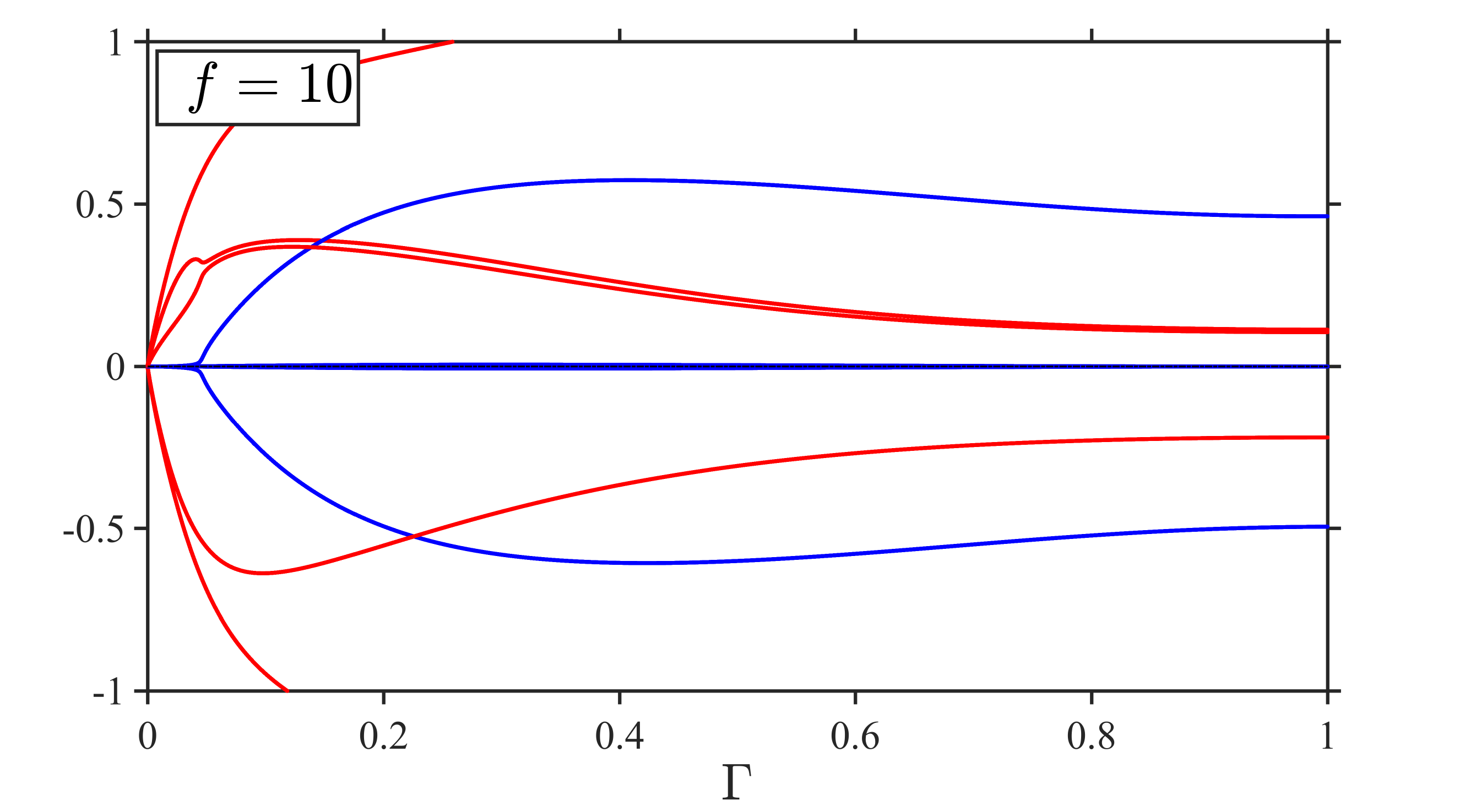}
    \caption{Same panels as in Figure~\ref{Fig_inv} but for the viscous
    case with $\Ek=10^{-1}$, computed from the first term in expression 
    \rf{vis_BC_ell}.
    \label{Fig_vis}}
\end{figure}

Although the viscous formulation yields an explicit dispersion relation, the evaluation 
of the surface integrals entering \rf{Iint} remains technically challenging. Both integral 
contributions in \rf{vis_BC_ell} exhibit endpoint singularities and must be reformulated 
in terms of complete elliptic integrals of the first, second, and third kinds. However, 
the second contribution involves integrals that do not admit closed analytical forms and 
therefore require numerical quadrature. By contrast, the first viscous contribution can 
be evaluated explicitly (using symbolic calculus softwares like Maple or Mathematica), without 
resorting to numerical discretisation. We therefore restrict the present computations 
to the case where $\mathfrak{I}\equiv0$, which still captures a substantial viscous contribution. 
We emphasise that this restriction is motivated by analytical and numerical feasibility 
rather than asymptotic considerations, and that the neglected contribution is formally of the same 
order in $\Ek$.

The same set of panels as in Figure~\ref{Fig_inv} is shown in Figure~\ref{Fig_vis} for 
the viscous case with $\Ek = 10^{-1}$. Viscosity primarily manifests itself through the 
appearance of \textit{avoided crossings} in the neighbourhood of inviscid 
Hamilton--Hopf bifurcation points \citep{Labarbe}, while the global structure of the 
dispersion curves remains close to that of the inviscid case. Notably, we observe
viscosity-driven secular instabilities for Riemann ellipsoids outside the inviscid 
interval of instability. This behaviour slightly contrasts with that of Maclaurin spheroids, 
for which arbitrarily small viscosity destabilises the equilibrium at which the axisymmetric 
sequence bifurcates toward a triaxial configuration (namely the Jacobi ellipsoids). 
Riemann ellipsoids being intrinsically non-axisymmetric, viscosity does 
induce an analogous loss of stability but not through symmetry breaking. In the viscous 
framework, Riemann ellipsoids of almost all admissible shapes (except for the irrotational 
sequence with $f=-2$) are found to be unstable to second-order harmonic 
perturbations.

\section{Discussion} \label{sec_conclu}

In this work, we have developed a unified and tractable linear stability framework 
for S-type Riemann ellipsoids, addressing long-standing limitations of previous 
inviscid analyses \citep{C65,C66,LL96}. Earlier studies based on the virial method 
led to global tensorial equations whose complexity grows rapidly with the harmonic 
order, thereby restricting practical computations to second- and third-order normal 
modes \citep{C65}. As noted by \cite{LL96}, higher-order disturbances are expected 
to be less strongly influenced by the gravitational field. However, a systematic 
treatment of this regime had remained unavailable, except through short-wavelength 
approximations. By contrast, the present formulation provides a consistent local 
approach that is valid for arbitrary harmonic perturbations, while remaining 
computationally efficient across the relevant parameter space. 

In the inviscid limit, we derived a generalised Poincar\'e-type equation governing 
small-amplitude oscillations of a triaxial ellipsoid with internal strain. We showed 
that its solutions form a one-parameter family of polynomials, extending the 
classical results of \cite{P85} and \cite{C22}, which are recovered as special 
cases. This formulation offers a systematic alternative to the virial tensor method 
traditionally employed for Riemann ellipsoids \citep{C65,C66}, while avoiding the 
wavelength restrictions inherent to short-wavelength (WKB) analyses \citep{LL96}. 
As a result, the full inviscid spectrum can be computed explicitly for arbitrary 
ellipsoidal harmonics, yielding dispersion relations in closed-form. This represents 
an extension of earlier investigations of differentially rotating self-gravitating 
bodies \citep{LBO67}. Notably, we show that almost all admissible S-type Riemann 
ellipsoids are unstable to second-order harmonic disturbances.

Building on these results, we examined the influence of weak viscosity 
by developing a small-Ekman-number asymptotic theory based on Prandtl’s 
boundary-layer framework \citep{LandauLifshitz}. For the case $n=2$, where the 
inviscid velocity field remains harmonic, viscous effects enter at leading order 
solely through the surface stress conditions. By resolving the associated 
boundary-layer dynamics analytically, we derived first-order viscous corrections 
to the inviscid dispersion relation and obtained an explicit criterion for secular 
instability. This analysis provides a systematic description of dissipation-induced 
instabilities in Riemann ellipsoids, closely following the classical treatments 
available for Maclaurin spheroids \citep{RS63,C79,Labarbe}. As in those systems, 
dissipation acts as a destabilising mechanism \citep{BKMR94}, undermining the 
gyroscopic balance that sustains rotating self-gravitating configurations 
\citep{KelvinI}. However, a definitive assessment of viscous stability requires 
the numerical evaluation of the second contribution in \rf{vis_BC_ell}, which is 
formally of the same asymptotic order but involves surface integrals that do not 
admit closed-form expressions. The development of robust numerical methods to compute 
this term is left for future work.

A natural extension of this work concerns the Chandrasekhar--Friedman--Schutz 
instability \citep{FS78a,FS78b}. Addressing this problem requires incorporating 
relativistic corrections to the Newtonian gravitational potential, following the 
pioneering approach of \cite{C70}. The destabilisation of rapidly rotating 
self-gravitating stars by gravitational radiation reaction remains a compelling 
target for future gravitational-wave observations. Although direct detection from 
isolated stellar objects remains beyond the sensitivity of current interferometers 
\citep{A16}, extending existing analyses \citep{Labarbe} to include internal fluid 
motions may contribute to improved modelling and parameterisation for next-generation 
detectors.

Beyond their mathematical interest, the results we presented bear direct relevance 
for geophysical and astrophysical fluid dynamics. Riemann ellipsoids serve as 
idealised yet physically meaningful models of rotating planets, stars, and fluid 
interiors with internal strain. Their stability properties play a central role in 
secular evolution, mode selection, and energy dissipation in such systems. The 
framework developed in this study opens the door for systematic investigations of 
linear instabilities in broader classes of rotating self-gravitating flows. \\


\noindent{\bf Funding.} This work was supported by the French government through 
the France 2030 investment plan managed by the National Research Agency (ANR), as 
part of the Initiative of Excellence Universit\'e C{\^o}te d’Azur under reference
number ANR-15-IDEX-01. \\

\noindent{\bf Declaration of interests.} The author reports no conflict of interest.


\appendix

\section{Description of the ellipsoidal system of coordinates} \label{app1}

Cartesian coordinates $(x,y,z)$ are expressed in terms of the confocal ellipsoidal 
coordinates $(\lambda,\mu,\varphi)$ through the mapping
\begin{equation}
\label{xyz}
x^2 = \frac{\lambda^2\mu^2\varphi^2}{d^2_\mu d^2_\varphi} , \;\;
y^2 = \frac{(\lambda^2-d_\varphi^2)(\mu^2-d_\varphi^2)(d_\varphi^2-\varphi^2)}{d^2_\lambda d^2_\varphi} , \;\;
z^2 = \frac{(\lambda^2-d_\mu^2)(d_\mu^2-\mu^2)(d_\mu^2-\varphi^2)}{d^2_\lambda d^2_\mu} ,
\end{equation}
where the positive semifocal distances are defined by
\begin{equation}
d_\lambda \eqdef \sqrt{\Gamma^2-\Xi^2} , \;\;
d_\mu \eqdef \sqrt{1 - \Xi^2} , \;\;
d_\varphi \eqdef \sqrt{1 - \Gamma^2} .
\end{equation}

The geometric surfaces defined by constant values of $(\lambda,\mu,\varphi)$ 
correspond, respectively, to ellipsoids, one-sheet hyperboloids, and two-sheet 
hyperboloids. The surface of the unperturbed ellipsoid --- defined by the equilibrium 
condition \rf{ellipsoid} --- is obtained by setting $\lambda = 1$. Accordingly, the 
fluid domain occupied by a steady Riemann ellipsoid is given by
\[
\mathcal{D}_0 \eqdef \big\{ (\lambda,\mu,\varphi) : 
d_\mu \leq \lambda \leq 1 , \;
d_\varphi \leq \mu \leq d_\mu , \;
-d_\varphi \leq \varphi \leq d_\varphi \big\} .
\]
The associated orthonormal basis $(\bm{e}_\lambda,\bm{e}_\mu,\bm{e}_\varphi)$ is 
particularly convenient for boundary-value problems, since the outward unit 
normal vector at the free surface is simply $\bm{n} \eqdef \bm{e}_\lambda$.

Because the coordinate system is orthogonal, the metric scale factors are given 
by the square roots of the diagonal metric components, namely
\begin{equation}
h^2_\lambda \eqdef \frac{(\lambda^2-\mu^2)(\lambda^2-\varphi^2)}{(\lambda^2-d^2_\mu)(\lambda^2-d^2_\varphi)} , \quad
h^2_\mu \eqdef \frac{(\mu^2-\lambda^2)(\mu^2-\varphi^2)}{(\mu^2-d^2_\mu)(\mu^2-d^2_\varphi)} , \quad
h^2_\varphi \eqdef \frac{(\varphi^2-\lambda^2)(\varphi^2-\mu^2)}{(\varphi^2-d^2_\mu)(\varphi^2-d^2_\varphi)} .
\end{equation}

For axisymmetric configurations, it is often advantageous to work instead with 
oblate or prolate spheroidal coordinates. The correspondence between these 
coordinate systems and the general ellipsoidal geometry provides valuable 
simplifications in symmetry-reduced problems, as discussed in detail in the 
monograph by \cite{Dassios} and references therein.

\section{Geometric relations in curvilinear coordinates} \label{app2}

Let $\bm{f}$ be an arbitrary vector in $\mathbb{R}^d$. Its components in two 
coordinate systems $\{u_i\}_{i=1,\dots,d}$ and $\{v_i\}_{i=1,\dots,d}$ are denoted by 
$\{f_{u}\}_i$ and $\{f_{v}\}_i$, respectively. The transformation of vector 
components between these two bases is given by \citep{MorseFeshbach}
\begin{equation}
\label{vec_conv}
f_{v_i} = \frac{1}{h_{v_i}} \sum_{j=1}^{d} \frac{\partial u_j}{\partial v_i} f_{u_j} ,
\;\; i=1,\dots,d ,
\end{equation}
where $h_{v_i}$ denotes the metric scale factor associated with the coordinate 
$v_i$.

Let $T$ be a tensor of rank 2 whose components $T_{ij}$ are expressed in the 
first coordinate basis, as defined previously. The covariant transformation 
of this tensor on the second orthonormal curvilinear basis is given by \citep{MorseFeshbach}
\begin{equation}
\label{tensor_ell}
T_{v_iv_j} = \frac{1}{h_{v_i}h_{v_j}} \sum_{k=1}^{d} \sum_{l=1}^{d}
\frac{\partial u_k}{\partial v_i} \frac{\partial u_l}{\partial v_j} T_{u_ku_l} 
\;\; i=1,\dots,d ,\;\; j=1,\dots,d ,.
\end{equation}

Accordingly, the gradient and Laplacian operators in an orthogonal curvilinear 
coordinate system $\{u_i\}_{i=1}^d$ take the standard form \citep{MorseFeshbach}
\begin{equation}
\label{del_ell}
\nab \eqdef \sum_{j=1}^{d} \frac{\bm{e}_{u_j}}{h_{u_j}} \frac{\partial}{\partial u_j} , \;\;
\nab^2 \eqdef J^{-1} \sum_{j=1}^{d} \frac{\partial}{\partial u_j} \left( \frac{J}{h_{u_j}^2} 
\frac{\partial}{\partial u_j} \right) ,
\end{equation}
where $J\eqdef \prod_{j=1}^{d} h_{u_j}$ is the Jacobian determinant.

\section{Definition of ellipsoidal harmonics} \label{app3}

Ellipsoidal harmonics are the Lam\'e eigenfunctions of the Laplace operator 
expressed in ellipsoidal coordinates, see \rf{del_ell}. They form an orthogonal 
basis and provide a natural framework for representing sufficiently regular 
scalar fields in ellipsoidal domains \citep{Dassios}.

We denote by $E_n^m(x)$ the ellipsoidal harmonic of degree $n$ and order $m$ of 
the first kind, associated with the coordinate $x$ (which may represent 
$\lambda$, $\mu$, or $\varphi$). These functions belong to one of four classical 
families of Lam\'e functions, given by
\begin{align}
K_n^m(x) &\eqdef \sum_{k=0}^{r} \widehat{a}_k (p^{K}_m) x^{n-2k} , \label{Knm} \\
L_n^m(x) &\eqdef \sqrt{x^2-d_\varphi^2} \sum_{k=0}^{r-1} \widehat{b}_k (p^{L}_m) x^{n-1-2k} , \label{Lnm} \\
M_n^m(x) &\eqdef \sqrt{x^2-d_\mu^2} \sum_{k=0}^{r-1} \widehat{b}_k (p^{M}_m) x^{n-1-2k} , \label{Mnm} \\
N_n^m(x) &\eqdef \sqrt{x^2-d_\varphi^2} \sqrt{x^2-d_\mu^2} \sum_{k=0}^{r-1} \widehat{c}_k (p^{N}_m) x^{n-2-2k} , \label{Nnm}
\end{align}
where $r \eqdef \lfloor n/2 \rfloor$ and $p_m$ denote the roots of the associated 
characteristic polynomials.

These characteristic polynomials $\mathcal{P}_m(p)$ arise from tridiagonal matrix 
eigenvalue problems and satisfy the recurrence relation
\begin{equation}
\pazocal{P}_{m} (p) \eqdef \bm{v}^{(d)}(p) \pazocal{P}_{m-1}(p) -
\bm{v}^{(u)} \bm{v}^{(l)} \pazocal{P}_{m-2}(p) ,
\end{equation}
with $\pazocal{P}_{-1} \eqdef 0$, $\pazocal{P}_{0} \eqdef 1$. 
The vectors $\bm{v}^{(d)}$, $\bm{v}^{(u)}$, and $\bm{v}^{(l)}$ correspond to the diagonal, 
upper, and lower bands of the tridiagonal matrix, respectively.

For ellipsoidal harmonics, these coefficients take the explicit forms given in 
\cite{Dassios}, which we reproduce here for completeness. The roots $p_m^{K}$ 
are obtained from
\begin{alignat*}{2}
v^{(d)}_{i}(p) \eqdef -(d_\mu^2 + d_\varphi^2) [ p - \left( n-2i+2 \right)^2 ],
\; i&=1,\dots,r+1 , \\
v^{(u)}_{i}(p) \eqdef 2i (2n-2i+1), \;
v^{(l)}_{i}(p) \eqdef - d_\mu^2 d_\varphi^2 (n-2i+1) (n-2i+2),
\; i&=1,\dots,r .
\end{alignat*}
The roots $p_m^{L}=p_m^{M}$ are determined from
\begin{alignat*}{2}
v^{(d)}_{i}(p) \eqdef -(d_\mu^2 + d_\varphi^2) [ p - \left( n-2i+1 \right)^2 ] +
h_\mu^2 (2n-4i+3), \; i&=1,\dots,r+1 , \\
v^{(u)}_{i}(p) \eqdef 2i (2n-2i+1), \;
v^{(l)}_{i}(p) \eqdef -d_\mu^2 d_\varphi^2 (n-2i+1) (n-2i),
\; i&=1,\dots,r-1 ,
\end{alignat*}
and the roots $p_m^{N}$ from
\begin{alignat*}{2}
v^{(d)}_{i}(p) \eqdef -(d_\mu^2 + d_\varphi^2) [ p - \left( n-2i+1 \right)^2 ], 
\; i&=1,\dots,r , \\
v^{(u)}_{i}(p) \eqdef 2i (2n-2i+1), \;
v^{(l)}_{i}(p) \eqdef -d_\mu^2 d_\varphi^2 (n-2i-1) (n-2i),
\; i&=1,\dots,r-1 .
\end{alignat*}

The expansion coefficients $(\widehat{a},\widehat{b},\widehat{c})$ appearing in 
\rf{Knm}--\rf{Nnm} are obtained from recurrence relations. For $K_n^m$ and $k\ge2$,
\begin{equation*}
\widehat{a}_{k} \eqdef \frac{(d_\mu^2 + d_\varphi^2) [ p - \left( n-2k+2 \right)^2 ] \widehat{a}_{k-1} + 
d_\mu^2 d_\varphi^2 (n-2k+3)(n-2k+4) \widehat{a}_{k-2}}{2k(2n-2k+1)} ,
\end{equation*}
with $\widehat{a}_0\eqdef 1$ and $\widehat{a}_1\eqdef(d_\mu^2+d_\varphi^2)(p-n^2)/[2(2n-1)]$.
For the Lam\'e functions $L_n^m$ and $M_n^m$ --- since their expansion coefficients are the same --- 
and $k\geq3$, it yields
\begin{align*}
\widehat{b}_{k-1} \eqdef &\frac{(d_\mu^2 + d_\varphi^2) [ p - \left( n-2k+3 \right)^2 ] - d_\mu^2(2n-4k+7)}{2(k-1)(2n-2k+3)} 
\widehat{b}_{k-2} \\
&+ \frac{d_\mu^2 d_\varphi^2 (n-2k+5) (n-2k+4)}{2(k-1)(2n-2k+3)} \widehat{b}_{k-3} ,
\end{align*}
with $\widehat{b}_0\eqdef 1$ and $\widehat{b}_1\eqdef[(d_\mu^2+d_\varphi^2)(p-(n-1)^2)-d_\mu^2(2n-1)]/[2(2n-1)]$.
For the Lam\'e function $N_n^m$ and $k\geq3$, it yields
\begin{equation*}
\widehat{c}_{k-1} \eqdef \frac{(d_\mu^2 + d_\varphi^2) [ p - \left( n-2k+3 \right)^2 ] \widehat{c}_{k-2} + 
d_\mu^2 d_\varphi^2 (n-2k+3)(n-2k+4) \widehat{c}_{k-3}}{2(k-2)(2n-2k+3)} ,
\end{equation*}
with $\widehat{c}_0\eqdef 1$ and $\widehat{c}_1\eqdef (d_\mu^2+d_\varphi^2)[p-(n-1)^2]/[2(2n-1)]$.
Efficient numerical strategies for computing the roots and coefficients are discussed extensively in \cite{Dassios}.

The ellipsoidal harmonics of the second kind, also known as exterior solutions, 
are defined by
\begin{equation}
F_n^m(x) \eqdef (2n+1) E_n^m(x) \int_{x}^{\infty} \frac{\ud s}{[E_n^m(s)]^2 
(s^2 - d_\mu^2)^{1/2} (s^2 - d_\varphi^2)^{1/2}} .
\end{equation}

Ellipsoidal harmonics provide a natural basis for expanding square-integrable 
functions on the ellipsoid. For any $f\in L^2(\mathcal{D}_0)$, we write
\begin{equation}
\label{ell_expand}
f(\lambda,\mu,\varphi,t) = \sum_{n=0}^{\infty} \sum_{m=0}^{2n+1} \left[ \widehat{f}^{(i)}_{nm} (t) E^n_m (\lambda) +
\widehat{f}^{(e)}_{nm} (t) F^n_m (\lambda) \right] \mathbb{S}_n^m (\mu,\varphi) ,
\end{equation}
where $\mathbb{S}_n^m(\mu,\varphi)\eqdef E_n^m(\mu)E_n^m(\varphi)$ denote the surface 
ellipsoidal harmonics.

These functions form a complete orthogonal set with respect to the surface 
measure
\begin{equation}
\label{surf_elem}
\ud S \eqdef w h_\mu h_\varphi \ud \mu \ud \varphi , \quad \textrm{with} \quad
w \eqdef \frac{1}{\sqrt{(1-\mu^2)(1-\varphi^2)}} ,
\end{equation}
and satisfy the orthogonality relation
\begin{equation}
\int_{\mathcal{D}_0} \mathbb{S}_n^m(\mu,\varphi) \mathbb{S}_q^p(\mu,\varphi) \ud S = 
\varpi_n^m \delta_{nq} \delta_{mp} ,
\end{equation}
with
\begin{equation}
\varpi_n^m = \int_{\mathcal{D}_0} \left[ \mathbb{S}_n^m(\mu,\varphi) \right]^2 \ud S .
\end{equation}

Finally, the Wronskian relation between interior and exterior harmonics reads
\begin{equation}
\label{Wronsk}
\pazocal{W}_n(\lambda) \eqdef F_n^m(\lambda) \frac{\partial}{\partial\lambda} E_n^m(\lambda)
- E_n^m(\lambda) \frac{\partial}{\partial\lambda} F_n^m(\lambda) 
= \frac{2n+1}{(\lambda^2-d_\mu^2)^{1/2}(\lambda^2-d_\varphi^2)^{1/2}} .
\end{equation}

\section{Coefficients of the integrated Prandtl equations} \label{app4}

We give the coefficients of the integrated Prandtl equations as
\begin{align*}
a_0 &= b_0 = - \varpi_n^m \Ro , \quad
a_1 = b_1 = \int \big[ h_\lambda u_{0\lambda} \big]_{\partial\mathcal{D}_0}
\big[ \mathbb{S}_n^m \big]^2 \ud S , \\
a_2 &= \int h_\lambda^2 \Big[ \Big( \frac{1}{h_\mu} \frac{\partial u_{0\mu}}{\partial\mu} 
- 2\ui\sigma\Ro \Big) + \frac{u_{0\mu}}{2 h_\mu} \frac{\partial}{\partial\mu} 
+ \frac{u_{0\varphi}}{2 h_\varphi} \frac{\partial}{\partial\varphi} 
\Big]_{\partial\mathcal{D}_0} \big[ \mathbb{S}_n^m \big]^2 \ud S , \\
b_2 &= \int h_\lambda^2 \Big[ \Big( \frac{1}{h_\varphi} \frac{\partial u_{0\varphi}}{\partial\varphi} 
- 2\ui\sigma\Ro \Big) + \frac{u_{0\mu}}{2 h_\mu} \frac{\partial}{\partial\mu} 
+ \frac{u_{0\varphi}}{2 h_\varphi} \frac{\partial}{\partial\varphi} 
\Big]_{\partial\mathcal{D}_0} \big[ \mathbb{S}_n^m \big]^2 \ud S , \\
a_3 &= \int h_\lambda \bigg[ \frac{h_\lambda}{h_\varphi} \frac{\partial u_{0\mu}}{\partial\varphi} 
- 2\Ro \frac{\partial z}{\partial\lambda} \bigg]_{\partial\mathcal{D}_0} \big[ \mathbb{S}_n^m \big]^2 \ud S , \quad
b_3 = \int h_\lambda \bigg[ \frac{h_\lambda}{h_\mu} \frac{\partial u_{0\varphi}}{\partial\mu}
+ 2\Ro \frac{\partial z}{\partial\lambda} \bigg]_{\partial\mathcal{D}_0} \big[ \mathbb{S}_n^m \big]^2 \ud S .
\end{align*}

Similarly, the coefficients of the tangent stress conditions are
\begin{equation*}
\tau_\mu = \int \bigg[
\frac{\partial}{\partial\lambda} \delta u_\mu^{(0)} + \frac{h_\lambda}{h_\mu}
\frac{\partial}{\partial\mu} \delta u_\lambda^{(0)} 
\bigg]_{\partial\mathcal{D}_0} \mathbb{S}_n^m \ud S , \quad
\tau_\varphi = \int \bigg[
\frac{\partial}{\partial\lambda} \delta u_\varphi^{(0)} + \frac{h_\lambda}{h_\varphi}
\frac{\partial}{\partial\varphi} \delta u_\lambda^{(0)} 
\bigg]_{\partial\mathcal{D}_0} \mathbb{S}_n^m \ud S .
\end{equation*}


\bibliographystyle{apacite}
\bibliography{Biblio}

\end{document}